\documentclass[aoas,MSNbibl,nameyear,dvips]{arximspdf}
\usepackage{graphicx}

\doi{10.1214/11-AOAS523}
\volume{6}
\issue{2}
\pubyear{2012}
\firstpage{697}
\lastpage{718}

\makeatletter
\def\bfA{\mathbf A}
\def\bfB{\mathbf B}
\def\bfI{\mathbf I}
\def\bfV{\mathbf V}
\def\bfw{\mathbf w}
\def\bfx{\mathbf x}
\def\bfy{\mathbf y}
\def\bfz{\mathbf z}
\def\bfbeta{\boldsymbol\beta}
\def\bfgamma{\boldsymbol\gamma}
\def\bfeta{\boldsymbol\eta}
\def\bfmu{\boldsymbol\mu}
\def\bfphi{\boldsymbol\phi}
\def\bfpsi{\boldsymbol\psi}
\def\bfLambda{\boldsymbol\Lambda}
\def\bfSigma{\boldsymbol\Sigma}
\newcommand{\vareps}{\varepsilon}

\makeatother

\begin{document}
\begin{frontmatter}

\title{Meta-analysis of functional neuroimaging data
using Bayesian nonparametric binary regression}
\runtitle{Bayesian meta-analysis of fMRI data}

\begin{aug}
\author[A]{\fnms{Yu Ryan} \snm{Yue}\corref{}\thanksref{au1}\ead[label=e1]{yu.yue@baruch.cuny.edu}},
\author[B]{\fnms{Martin A.} \snm{Lindquist}\thanksref{au2}\ead[label=e2]{martin@stat.columbia.edu}}
\and
\author[C]{\fnms{Ji Meng} \snm{Loh}\ead[label=e3]{loh@research.att.com}}
\runauthor{Y. R. Yue, M. A. Lindquist and J. M. Loh}
\affiliation{City University of New York, Columbia University and
AT\&T~Labs-Research}
\address[A]{Y. R. Yue\\
Department of Statistics and CIS\\
Zicklin School of Business\\
City University of New York\\
One Bernard Baruch Way\\
New York, New York 10010\\USA\\
\printead{e1}} %adresu isvedimo komanda gale!
\address[B]{M. A. Lindquist\\
Department of Statistics\\
Columbia University\\
1255 Amsterdam Ave \\
New York, New York 10027\\USA\\
\printead{e2}}
\address[C]{J. M. Loh\\
AT\&T Labs-Research\\
180 Park Ave-Building 103\\
Florham Park, New Jersey 07932\\USA\\
\printead{e3}}
\end{aug}
\thankstext{au1}{Supported by a PSC-CUNY Award,
jointly funded by the Professional Staff Congress and the City
University of New York.}
\thankstext{au2}{Supported in part by NSF Grant DMS-08-06088.}

% HISTORY:
\received{\smonth{1} \syear{2011}}
\revised{\smonth{10} \syear{2011}}

% ABSTRACT
%
\begin{abstract}
In this work we perform a meta-analysis of neuroimaging data,
consisting of locations of peak activations identified in $162$
separate studies on emotion. Neuroimaging meta-analyses are typically
performed using kernel-based methods. However,
these methods require the width of the kernel to be set a priori and to
be constant across the brain. To address
these issues, we propose a fully Bayesian nonparametric binary
regression method to perform neuroimaging meta-analyses.
In our method, each location (or voxel) has a probability of being a
peak activation, and the corresponding probability
function is based on a spatially adaptive Gaussian Markov random field
(GMRF). We also include parameters in the model to
robustify the procedure against miscoding of the voxel response.
Posterior inference is implemented using efficient MCMC
algorithms extended from those introduced in Holmes and Held
[\textit{Bayesian Anal.}
\textbf{1} (2006)
145--168]. Our
method allows the probability function to be
locally adaptive with respect to the covariates, that is, to be smooth
in one region of the covariate space and wiggly
or even discontinuous in another. Posterior miscoding probabilities for
each of the identified voxels can also be obtained,
identifying voxels that may have been falsely classified as being
activated. Simulation studies and application to the emotion
neuroimaging data indicate that our method is superior to standard
kernel-based methods.
\end{abstract}

% KEYWORDS
%

\end{frontmatter}

%s1 #&#
\section{Introduction}\label{sec1} \label{sectintro}
%{equation}{0}
%s1.1 #&#
\subsection{Meta-analysis of neuroimaging studies}\label{sec1.1}
In recent years there has been a rapid increase in the number and
variety of neuroimaging studies being\vadjust{\goodbreak} performed around the world. This
growing body of knowledge is accompanied by a need to integrate
research findings and establish consistency across labs and scanning
procedures, and to identify consistently activated regions across a set
of studies. Performing meta-analyses has become the primary research
tool for accomplishing this goal [\citet{WagerSCAN2007};
\citet{WagerNI2009}].
Evaluating consistency is important because false positive rates in
neuroimaging studies are likely to be higher than in many fields, as
many studies do not adequately correct for multiple comparisons. Thus,
some of the reported activated locations are likely to be false
positives, and it is important to assess which findings have been
replicated and have a higher probability of being real activations.
Individual imaging studies often use very different analyses [see
\citet
{Lind2008} for an overview], and effect sizes are only reported for a
small number of activated locations, making combined effect-size maps
across the brain impossible to reconstruct from published reports.
Instead, meta-analysis is typically performed on the spatial
coordinates of peaks of activation (peak coordinates), reported in the
standard coordinate systems of the Montreal Neurologic Institute (MNI)
or \citet{TalTour88}, and combined across studies. These peak
coordinates typically correspond to the voxel whose $t$-statistic takes
the maximum value in a spatially coherent cluster of activation, that
is, the max statistic among a~set of adjacent voxels that exceed a~certain threshold. This information is typically provided in most
neuroimaging papers and simple transformations between the two standard
spaces exist.

A typical neuroimaging meta-analysis studies the locations of peak
activations from a large number of studies and seeks to identify
regions of consistent activation. This is usually performed using
kernel-based methods such as activation likelihood estimation [ALE;
\citet{Turk2002}] or kernel density approximation [KDA;
\citet
{Wager2004}]. In both methods, maps are created for each study by
convolving an indicator map, consisting of an impulse response at each
study peak, with a kernel of predetermined shape and width. The
resulting maps are thereafter combined across studies to create a
meta-analysis map. Monte Carlo methods are used to find an appropriate
threshold to test the null hypothesis that the $n$ reported peak
coordinates are uniformly distributed throughout the grey matter. A
permutation distribution is computed by repeatedly generating $n$ peaks
at random locations and performing the smoothing operation to obtain a
series of statistical maps under the null hypothesis that can be used
to compute voxel-wise $p$-values. The two approaches differ in the
shape of the smoothing kernel. In KDA, it is assumed to be a sphere
with fixed radius, while in ALE it is a Gaussian with fixed standard deviation.

A major shortcoming of kernel-based approaches is that the width of the
kernel, and thus the amount of smoothing, is fixed  a priori
and assumed to be constant throughout the brain. In order to address
these concerns, we propose a fully Bayesian nonparametric binary
regression method for performing neuroimaging meta-analysis. In our
method, each location has a probability of being a peak activation, and
the corresponding probability function is based on a spatially adaptive
Gaussian Markov random field (GMRF). The locally adaptive features of
our method allows us to better match the natural spatial resolution of
the data across the brain compared to using an arbitrary chosen fixed
kernel size.

%
%f1 #&#
\begin{figure}

\includegraphics{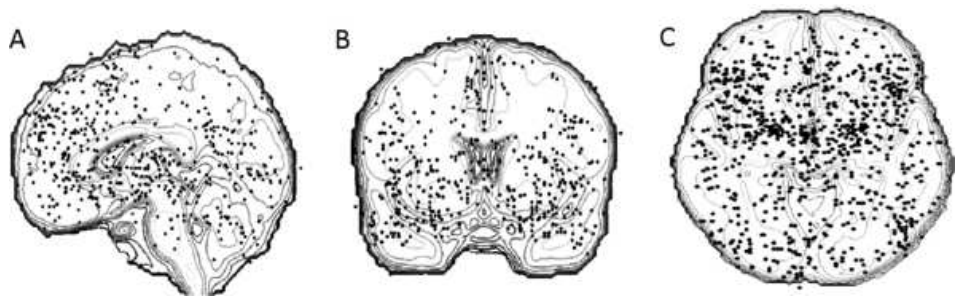}

\caption{Example of the raw data are shown for a representative
sagittal, coronal and axial slice of the brain. Each point represents a
reported activation foci in an individual study by criteria designated
by that particular study. All foci are reported and plotted in the MNI
brain template to allow for cross study comparisons.}
\label{figdata}
\end{figure}

In this work, a meta-analysis was performed on the results of $162$
neuroimaging studies ($57$ PET and $105$ fMRI) on emotion. The studies
were all performed on healthy adults and published between 1990 and
2005. For each study, the foci of activation were included when
reported as significant by the criteria designated in the individual
studies. Relative decreases in activation in emotion related tasks were
not analyzed. All coordinates were reported on the MNI coordinate
system to allow for cross study comparisons. Together, these studies
yield a data set consisting of $2478$ unique peak coordinates. This
data set is described in greater detail in \citet{Kober2008}. Due
to the
relative scarcity of neuroimaging studies on a particular topic (e.g.,
emotion), it is standard practice in meta-analysis to combine data
obtained using different imaging modalities, sample sizes and
statistical analyses. This is done to ensure that the analysis has
enough power to detect effects of interest. In addition, studies in
\citet{WagerHB2008} have shown no significant difference between
MRI and
PET in the assessment of their functional maps and their foci of
activation. Figure \ref{figdata} shows the raw data for representative
slices of the brain with fixed $x$, $y$ and $z$ directions,
respectively. Each point in the plot represents the location of the
peak of a cluster of reported activation from one of the $162$
neuroimaging studies. The primary goal for analyzing this data set was
to determine areas of the brain that are consistently active in studies
of emotion.

%s1.2 #&#
\subsection{Statistical modeling for binary response}\label{sec1.2}

%We propose a Bayesian nonparametric binary regression method to
%analyze the data described above.
Let $Y$ be a random binary response variable, $X$ a vector of
covariates and $p(\bfx)$ the response probability function, $p(\bfx) =
\operatorname{Pr}(Y=1| X=\bfx)$. In the context of fMRI meta-analysis, $Y=1$
if the voxel is reported as being a peak activation. The vector $X$
includes the voxel location and possibly other covariates related to
the patient or the study. In nonparametric binary regression, we have
$p(\bfx) = H(z(\bfx))$, where $H$ is a specified cumulative
distribution function often referred to as the link function. Popular
link functions are the standard logistic and standard normal cumulative
density functions.

The traditional
parametric approach to binary regression involves setting $z(\bfx) =
\alpha+ \bfbeta^T\bfx$, with unknown parameters $\alpha$ and
$\bfbeta
$. \citet{Mccullagh1989} contains a comprehensive treatment of
frequentist parametric methods with exponential family models, binary
regression being a special case. Bayesian binary regression is well
documented in, for example,
\citet{deyghomal99}. In particular, \citet{albechib93} and
\citet
{holmheld06} introduced auxiliary variable methods that provide
efficient Markov chain Monte Carlo (MCMC) inference for parametric
binary regression.

There is an extensive non-Bayesian literature on nonparametric
regression using exponential family models, with binary regression
treated as a special case. \citet{OSullivan86} estimated a single
function using a penalized likelihood approach, and their work was
extended to additive models by \citet{Hastie90}. \citet
{gu90} and \citet
{wahba97} used tensor product smoothing splines to allow for
interactions between variables and estimated smoothing parameters via a
generalized cross-validation technique. \citet{load99} proposed a local
likelihood approach for both univariate and bivariate nonparametric
estimation and provided data-driven bandwidth estimators.

Bayesian methods for nonparametric
binary regression were developed in
\citet{woodkohn98}, \citet{holmmal03}, \citet
{choughoroy07}, and
\citet{trimulspl09}. These methods are not locally
adaptive, however. \citet{KrivCraiKauefast2008}
proposed an adaptive penalized spline estimator for binary regression
based on quasi-likelihoods. \citet{woodkohn08} presented a
locally adaptive Bayesian estimator for binary regression by using a
mixture of probit regressions where the argument of each probit
regression is a thin-plate spline prior with its own smoothing
parameters and the mixture weights depend on the
covariates.
%spline
%estimator with a Normal-Exponential-Gamma
%prior, which can be used to obtain adaptive estimates for
%one-dimensional binary regression only. Their method was not extended
%to
%the spatial smoothing scenario, however.

In fMRI meta-analysis, kernel-based smoothing techniques are typically
used to identify regions of consistent activation and Monte-Carlo
procedures are used to establish statistical significance. These
techniques count the
number of activation peaks within a radius of each local brain area
and compare the observed number to a null distribution to
establish significance. The kernel radius is chosen by the analyst,
and kernels that match the natural spatial resolution of the data are
the most statistically powerful [\citet{WagerSCAN2007}].
In our method, the function $z(\cdot)$ is assumed to be a spatially
adaptive Gaussian Markov random field (GMRF) with locally varying
variance. The local adaptiveness of the procedure allows
the probability function to be smooth in some regions and wiggly in
others, depending on the data information. The need of adaptive
smoothing for fMRI data has been demonstrated in \citet
{BrezgerFahmhen2007} and \citet{yuelohlindfmri2009}. The proposed
Bayesian nonparametric binary regression method is an extension to the
binary response case of methods developed in \citet
{yuespecknon09} and
\citet{yuelohpp10}. To make this procedure better suited for application
to fMRI meta-analysis, we incorporate additional model parameters
associated with the probabilities of voxels being miscoded. This makes
the modeling more robust to possible errors in the data. The posterior
inference is carried by efficient MCMC algorithms extended from those
in \citet{holmheld06}. From the model fit we obtain a map of the
probability of observing a peak activation across the brain as well as
posterior miscoding probabilities. Regions of the brain with high
probability estimates are identified as activated based on the
meta-analysis. This makes the proposed method far more interpretable
than earlier approaches.

The rest of the paper is organized as follows. The proposed
method is described in Section \ref{sectmodel}. Section
\ref{sectsimstudy} presents simulation studies comparing our
method to other available methods. Results of the data
analysis are given in Section \ref{sectanalysis}. Section \ref{sectsummary}
concludes this work with discussions.

%s2 #&#
\section{Bayesian hierarchical modeling and inference}\label{sec2} \label{sectmodel}

We describe in this section our nonparametric binary regression model
using the spatially adaptive GMRF. Note that our method currently can
only be implemented in two dimensions. We apply it to the fMRI setting
by fitting the model to brain slices in succession. This is similar to
the staggered approach in \citet{penny05}, who used a two-dimensional
Laplacian prior that is related to our GMRF prior.

%Section \ref{subsecspatialadaptive} describes a spatially adaptive
%GMRF on a two-dimensional regular lattice as developed in
%In Section \ref{subsecmcmc} we derive MCMC algorithms that allow us to
%perform
%posterior inference for binary regression using spatially adaptive
%GMRFs. Our robustification procedure is described in Section
%
%s2.1 #&#
\subsection{Spatially adaptive GMRF on regular lattice}\label{sec2.1} \label
{subsecspatialadaptive}
%In this section, we are going to introduce the GMRF prior which
%embodies our prior knowledge that evoked responses are spatially
%contiguous. This spatial prior is related to the Laplacian prior used
%in \citet{penny05} and currently only works for two-dimensional
%data
%inputs.

Let us denote by $\bfx= (x_{11}, x_{21}, \ldots, x_{n_1,n_2})'$ an
$n$-dimensional vector of voxel locations on a regular $n_1\times n_2$
lattice ($n=n_1n_2$). Adopting notation $z_{jk} = z(x_{jk})$, we assume
that the underlying spatial process $z_{jk}$ is an adaptive Gaussian
Markov random field (GMRF) as introduced in \citet{yuespecknon09}.
%{\bf This GMRF is derived by approximating a two-dimensional
%thin-plate spline (TPS) estimator
%both large-scale and local spatial dependence among those voxels. We
%will briefly describe the adaptive GMRF in this section and refer the
%reader to \citet{yue:speck:non:09} for details.}
This adaptive GMRF is based on the following spatial Gaussian random
walk model:
%
%e1 #&#
\begin{eqnarray}\label{adgmrf}
\bigl(\nabla^2_{(1,0)}+\nabla^2_{(0,1)} \bigr)z_{jk}\sim\mathrm{N}
(0,\delta^2\gamma_{jk}^2 ),
\end{eqnarray}
where $\nabla^2_{(1,0)}$ and $\nabla^2_{(0,1)}$ denote the second-order
backward difference operators in the vertical and horizontal directions
respectively, that is, $\nabla^2_{(1,0)}z_{jk}= z_{j+1,k}-2z_{jk} +
z_{j-1,k}$ and $\nabla^2_{(0,1)}z_{jk}=z_{j,k+1}-2z_{jk}+z_{j,k-1}$ for
$2\le j \le n_1-1$ and $2\le j \le n_2-1$. The parameter $\delta^2$ is
a global smoothing parameter accounting for large-scale spatial
variation while $\gamma_{jk}^2$ are the adaptive smoothing parameters
that capture the local structure of the process $z(\bfx)$. The equation
(\ref{adgmrf}) essentially defines an adaptive smoothness prior on the
second-order difference
$(\nabla^2_{(1,0)}+\nabla^2_{(0,1)})z_{jk}$. As a result, the
conditional distribution of each $z_{jk}$ given the rest $\bfz_{-jk}$
is Gaussian and only depends on
its neighbors in a specific way. This dependence can be shown using a
graphical notation by expressing the conditional expectation of an
interior $z_{jk}$ as
%
%e2 #&#
\begin{eqnarray}\label{expectzcond}
E(z_{jk}|\bfz_{-jk})=\frac{1}{20} \left(
8
\begin{array}{ccccc}
\circ&\circ&\circ&\circ&\circ\\[-6pt]
\circ&\circ&\bullet&\circ&\circ\\[-6pt]
\circ&\bullet&\circ&\bullet&\circ\\[-6pt]
\circ&\circ&\bullet&\circ&\circ\\[-6pt]
\circ&\circ&\circ&\circ&\circ
\end{array}
-2
\begin{array}{ccccc}
\circ&\circ&\circ&\circ&\circ\\[-6pt]
\circ&\bullet&\circ&\bullet&\circ\\[-6pt]
\circ&\circ&\circ&\circ&\circ\\[-6pt]
\circ&\bullet&\circ&\bullet&\circ\\[-6pt]
\circ&\circ&\circ&\circ&\circ
\end{array}
-1
\begin{array}{ccccc}
\circ&\circ&\bullet&\circ&\circ\\[-6pt]
\circ&\circ&\circ&\circ&\circ\\[-6pt]
\bullet&\circ&\circ&\circ&\bullet\\[-6pt]
\circ&\circ&\circ&\circ&\circ\\[-6pt]
\circ&\circ&\bullet&\circ&\circ
\end{array}
\right),
\end{eqnarray}
where the locations denoted by a ``$\bullet$'' represent those values
of $\bfz_{-jk}$ that the conditional expectation of $z_{jk}$
depends on, and the number in front of each grid denotes the weight
given to the corresponding ``$\bullet$'' locations. Therefore, the
conditional mean of $z_{jk}$ is a particular linear combination of
the values of its neighbors, and its conditional variance is $\operatorname
{Var} (z_{jk}|\bfz_{-jk})=20\delta^2\gamma_{jk}^2$.

The use of $\gamma_{jk}^2$ is important for estimating activation
probabilities in a~fMRI meta-analysis. To identify consistently
activated regions across a~set of studies, we need less smoothing
(large $\gamma_{jk}^2$) where there are many reported peak locations
and relatively more smoothing (small $\gamma_{jk}^2$) where very few or
no peaks are reported. Standard smoothing techniques (e.g., kernel
smoother with fixed width) suffer from a trade-off between increased
detectability and loss of information about the spatial extent and
shape of the activation areas. Adaptive smoothing provided by $\gamma
_{jk}^2$ can reduce such loss of information. The need of adaptive
smoothing for processing fMRI imaging data was also demonstrated in
\citet{BrezgerFahmhen2007} and \citet{yuelohlindfmri2009}.
Note that
setting $\gamma_{jk}^2\equiv1$ makes~(\ref{adgmrf}) a~nonadaptive GMRF
on lattice, yielding a Bayesian solution for thin-plate splines [see
\citet{GMRFbook}, section 3.4.2].

%with
%certain edge correction terms define an adaptive GMRF with matrix form
%given by
%where $\bfz=(z_{11},z_{21},\ldots,z_{n_1,n_2})'$, $\bfgamma=(
%$\bfA_\gamma=\bfB^T\operatorname{diag}(\bfgamma^{-1})\bfB$ is an $n\times n$ {
%full rank sparse matrix. The quantity $|\bfA_\gamma|_+$ is the product
%of the nonzero eigenvalues from $\bfA_\gamma$.
%that $|\bfA_
%easily computed. The prior (\ref{denadgmrf}) is an intrinsic GMRF
%because its density is improper Gaussian ($\bfA_\gamma$ is singular)
%and its attractive Markov property makes $\bfA_\gamma$ highly sparse:
%see \citet{yuespecknon09} for details.

Additional priors need to be specified for $\gamma_{jk}^2$ in (\ref
{adgmrf}). We use independent inverse gamma priors for $\gamma^2_{jk}$,
that is, $\gamma_{jk}^{-2}\stackrel{\mathrm{i.i.d.}}{\sim}\operatorname{Gamma}(\nu/2,1/2),
\nu>0$. The marginal prior distribution of the increment in (\ref
{adgmrf}) turns out to be a \mbox{Student-$t$} distribution with $\nu$ degrees
of freedom. We choose a Cauchy distribution ($\nu=1$), which has been
suggested as a default prior for
robust nonparametric regression [\citet{CartKohnmark1996}] and sparse
Bayesian learning [\citet{tippspar2001}]. \citet
{yuelohpp10} and \citet
{BrezgerFahmhen2007} also suggested similar priors for
$\gamma^2_{jk}$ in their work on adaptive spatial smoothing.
\citet
{yuespecknon09} and \citet{yuelohlindfmri2009}, however, assumed another
spatial GMRF model for $\log(\gamma^2_{jk})$
in a second hierarchy. Although it has been applied successfully for
modeling spatial data, this two-stage GMRF prior forces the $\gamma
_{jk}^2$ to be smooth and it is not suitable for estimating spatial
processes with jumps or sharp peaks. Furthermore, the computation is
rather complicated, precluding extensions to more flexible regression
models, for example, the binary hierarchical regression model
considered here.

The prior for $\delta^2$ is often chosen to be a conjugate diffuse but
proper inverse gamma prior. We, however, propose to use a half-$t$
distribution as the prior for its square root, that is,
%
%e3 #&#
\begin{eqnarray}\label{priorhcf}
[\delta|\rho,S]\propto\biggl(1 + \frac{1}{\rho} \biggl(\frac{\delta
}{S} \biggr)^2 \biggr)^{-(\rho+1)/2}, \qquad \delta> 0,
\end{eqnarray}
where $\rho$ is the parameter of degrees of freedom and $S$ is the
scale parameter. The half-$t$ distribution can be treated as the
absolute value of a Student-$t$ distribution centered at zero
[see \citet{psarpanft90}]. Although it is not commonly used in
statistics, the half-$t$ distribution was used
in objective Bayesian inference by \citet{wipegiropewshnt08} and
suggested for use as a default prior for a variance component in
hierarchical models [e.g., \citet{Gelmprio2006}; \citet{gelmanjcgs08}]. This
family includes, as special cases, the
improper uniform density (if $\rho=-1$) and the proper half-Cauchy (if
$\rho=1$). Following \citet{CarvPolsScothors2010}, we use a~standard
half-Cauchy prior ($\rho=S=1$) due to its heavy tail and substantial
mass around zero. Although it is not conjugate, the half-$t$ prior on
$\delta$ can be written as $\delta\stackrel{\mathcal{D}}{=}|\xi
|\theta
$, where $\xi\sim\mathrm{N}(0,1)$ and $\theta^2\sim\operatorname{IG}(\rho
/2,\rho
S^2/2)$ [e.g., \citet{psarpanft90}]. This property enables us to develop
efficient MCMC sampling schemes as shown in the \hyperref[appendix]{Appendix}.

%{\bf(Show the priors on other covariates if any)}

%s2.2 #&#
\subsection{Posterior inference}\label{sec2.2}%\label{subsecprobit}

Although any cumulative distribution function (cdf) $H$ that preserves
the smoothness of $\bfz$ may be used as a link function, here, we only
consider the case in which the $H$ can be represented as the scale
mixture of mean zero normal cdf's. Two special examples are the
well-known probit and logit link functions. With a specific link
function, the posterior distribution of $\bfz$ is not analytically
tractable, and thus an MCMC algorithm will be used to compute the
posterior distribution. The algorithm is based on the auxiliary
variable method in \citet{holmheld06} and GMRF simulation
techniques in
\citet{GMRFbook}. Briefly, the data are augmented by introducing an
auxiliary variable $w_i$ that follows a normal distribution with mean
$z_i$ and variance $\lambda_i$. The new data $w_i$ are associated with
original binary data $y_i$ in the following way: $y_i = 1$ if $w_i>0$
and $y_i = 0$ if $w_i\le0$. Then, the adaptive GMRF prior is taken on
$z_i$ and a certain prior distribution chosen for $\lambda_i$ depending
on the link function. The full conditional distributions for the Gibbs
sampler are all easily derived and can be efficiently sampled. In
the \hyperref[appendix]{Appendix} we provide the detailed MCMC algorithms for the
link functions that are the probit, logit and general scale mixture of normals.

%s2.3 #&#
\subsection{Robustification}\label{sec2.3} \label{subsecrobustification}
In this section we describe how to robustify our procedure against
miscoding of the response variable. Adopting the idea in
\citet{choughoroy07}, we use indicator variables
$\bfpsi=(\psi_1,\ldots,\allowbreak\psi_n)'$ such that $\psi_i=1$ indicates that
$y_i$ is miscoded and $\psi_i=0$ indicates that~$y_i$ is correctly
coded. In the context of fMRI meta-analysis, $\psi_i = 1$ means that
$y_i$ is either a false positive or a false negative. Since these
variables cannot be observed, we treat them as unknown parameters that
need to be estimated via taking priors on them. The joint posterior
distribution of $(\bfpsi, \bfz)$ is then
used to obtain a robust estimate of $\bfz$, and also to identify the
miscoded observations.

We assume that each observation has equal probability of being
miscoded, independent of other observations and $\bfz$. Denote by $r$
an  a priori  guess for the probability of an observation being
miscoded. Given $(\bfpsi, \bfz)$, the $y_i$'s are independent
Bernoulli random variables with probability of success
$(1-\psi_i)H(z_i) + \psi_i(1 - H(z_i))$. As a result, the conditional
distributions of $\psi_i$ are independent with
%
%e4 #&#
%e5 #&#
\begin{eqnarray}\label{fcpsi}
P(\psi_i=1|\bfy,\bfz)=
\cases{\displaystyle
\frac{r[1-H(z_i)]}{r[1-H(z_i)] + (1-r)H(z_i)}, &\quad  if  $y_i=1$,\vspace*{3pt}\cr\displaystyle
\frac{rH(z_i)}{rH(z_i) + (1-r)[1-H(z_i)]}, &\quad  if  $y_i=0$.}
\end{eqnarray}
Consider the probit link without any hyperprior. As shown in Section
\ref{subsecprobit}, we adjust latent variables $w_i$ for miscoding,
that is, $y_i=1$ if $\{w_i>0,\psi_i=0\}$ or $\{w_i<0,\psi_i=1\}$. Then,
%
%e6 #&#
%e7 #&#
%e8 #&#
\begin{eqnarray}\label{fcwrbst}
(w_i|\bfpsi,\xi,\bfeta,\bfy)\sim
\cases{\displaystyle
\mathrm{N}(\xi\eta_i,1)I(w_i>0), & \quad   if  $y_i+\psi_i=1$,\cr\displaystyle
\mathrm{N}(\xi\eta_i,1)I(w_i\le0), & \quad   if  $y_i+\psi_i\ne1$.}
\end{eqnarray}
Hence, samples from the joint distribution $(\psi_i,w_i|\bfz,\bfy)$
can be drawn by first sampling $\psi_i$ using (\ref{fcpsi}) and then
sampling $w_i$ using (\ref{fcwrbst}). Since the full conditional of
$\bfz$ does not depend on $\bfpsi$ or $\bfy$, the samples from the
conditional distributions of the rest of the parameters can be
drawn as described earlier. Note that the algorithm of this robust
approach may be extended similarly to the logit link or an arbitrary
symmetric link by introducing the relevant latent variables.

%
%f2 #&#
\begin{figure}
\begin{tabular}{cc}

\includegraphics{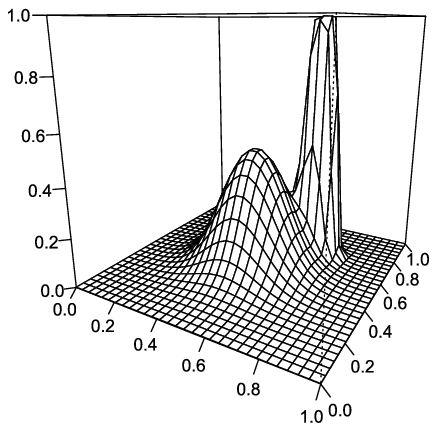}
& \includegraphics{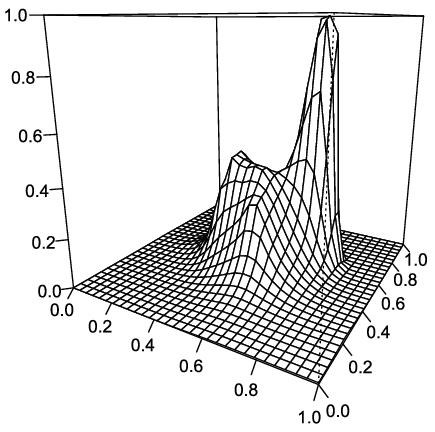}\\
(a)&(b)\\[6pt]

\includegraphics{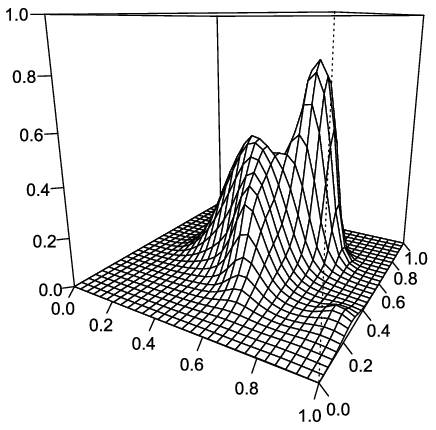}
&\includegraphics{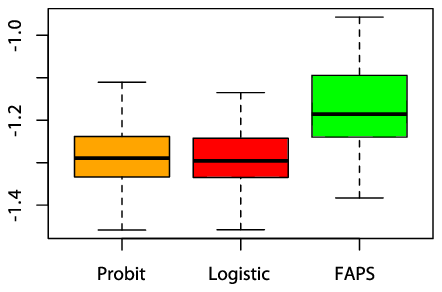}\\
(c)&(d)
\end{tabular}
\caption{Simulation I: \textup{(a)} True probability function;
\textup{(b)} Estimated
probability function using proposed method with probit link; \textup{(c)}
Estimated probability function using FAPS method; \textup{(d)} Mean squared
probability errors using proposed method and FAPS method.}
\label{figbimod}
\end{figure}

%Label & Function formula \\
%a (sine) & $p(x) = \Phi\{2\sin(4\pi x) \}$ \\
%b (peak) &$ p(x) = \Phi\{\fr{4}{3}\exp[-\fr{(x-0.6)^2}{0.18}
% ] +
% \fr{10}{3}\exp[-\fr{(x-0.1)^2}{0.001} ]-1.6 \}$\\
%c (step) & $ p(x)$ = $\Phi\{-1.036 + 2.073I_x(0.25) - 1.427
%I_x(0.75) \}$, $I_x(a)=1$ if $x>a$ \\
%d (bimodal) & $p(x,y)$ = $\Phi\{6\exp[-\fr{5}{2}
% ((x-2)^2+(y-2)^2 ) ]+3\exp[-\fr{1}{10}(x^2+y^2)
% ]- 3 \}$\\

%s3 #&#
\section{Simulation studies}\label{sec3}\label{sectsimstudy}
We performed two different types of simulation studies to investigate
the performance of our method. The first simulation is in the setting
of nonparametric binary regression,\vadjust{\goodbreak} where the proposed method is
compared to an adaptive penalized spline model. The second simulation
is in the setting of fMRI meta-analysis, where our method is compared
to the kernel-based ALE method, which is commonly used in neuroimaging
meta-analysis.

%s3.1 #&#
\subsection{Simulation I}\label{sec3.1} \label{subsectsim1}
The true probability function is assumed to be
\[
p(\bfx) = \Phi\biggl\{6\exp\biggl[-\frac{5}{2}
\bigl((x_1-2)^2+(x_2-2)^2 \bigr) \biggr]+3\exp\biggl[-\frac
{1}{10}(x^2_1+x_2^2) \biggr]- 3 \biggr\}.
\]
It is a smooth bimodal spatial surface on a $30\times30$ regular
lattice as shown in Figure \ref{figbimod}(a). One hundred data sets
were simulated and we use the mean squared probability error (MSPE),
\[
\mathrm{MSPE} = \frac{1}{n}\sum_{i=1}^n\{p(\bfx_{i}) - \hat{p}(\bfx
_{i})\}^2,
\]
to measure performance, where $\hat{p}(\cdot)$ is the estimated
probability function.

The estimates obtained using our Bayesian nonparametric binary
regression model are compared to those obtained using the fast adaptive
penalized splines (FAPS) model in \citet{KrivCraiKauefast2008}.
The FAPS
approach models the
regression function as a penalized spline with a smoothly varying
smoothing parameter function which is also modeled as a penalized
spline. Their method handles local smoothing of binary data as a
special case. The authors showed that the FAPS estimator outperformed
the penalized spline estimators in \citet{CraiRupp07} and
\citet{RuppCarrspat2000}. The model can be fit using
the \verb@AdaptFit R@ package.\looseness=-1

Panels (b) and (c) in Figure \ref{figbimod} show typical fits for the
bimodal function using our method and FAPS method, respectively. It
appears that the FAPS model has difficulty capturing the sharp peak and
undersmoothes the flat portion as well. Figure \ref{figbimod}(d) shows
the distributions of the MSPE produced by those two methods, where the
FAPS estimator is apparently outperformed. Also, in our method the two
link functions yield similar performances in terms of MSPE. This is
because nonparametric modeling of $\bfz$ makes the model robust against
the choice of the link function. We believe that the underperformance
of FAPS stems from using slowly varying functions to model local
smoothing parameters. Although they provide computational efficiency,
such low-rank basis functions are unable to capture sharp changes in
the function. \citet{yuespecknon09} presented similar results for normal
response variables. Note that the robustification procedure is not
required in this simulation study.\vspace*{-3pt}
%
%f3 #&#
\begin{figure}[b]
\vspace*{-3pt}
\includegraphics{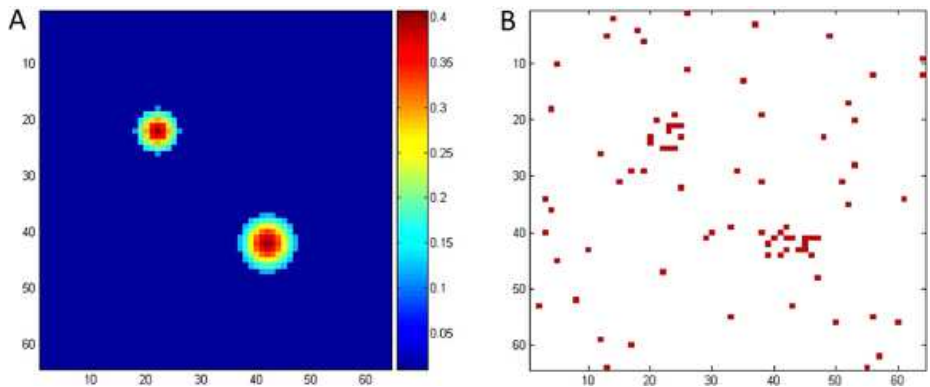}

\caption{\textup{(A)} The probability map used to generate random activation
peaks; \textup{(B)} One set of simulated activation peaks.}
\label{figsimdata}
\end{figure}

%s3.2 #&#
\subsection{Simulation II}\label{sec3.2} \label{subsectsim2}

In the second simulation study we began by constructing a $64 \times
64$ probability map,
denoted $p(x,y)$, where the value at each voxel location $(x,y)$
represents the probability that it be recorded as a~``peak coordinate''
in a neuroimaging study. The probability map consisted of two circular
regions of heightened probability (see Figure \ref{figsimdata}A),\vadjust{\goodbreak} where
the maximum probability is roughly $0.4$. Voxels lying outside these
two regions
were set to have a constant background probability of $0.01$, thus
allowing for the possibility of ``false positives'' outside the two
centers of activation. Next, the probability map was used
to generate random activation peaks. The voxel at coordinate $(x,y)$
was considered a reported
peak according to a binomial distribution with probability of
activation $p(x,y)$. This process
was repeated $100$ times and each time gives rise to simulated
meta-analysis data. Figure \ref{figsimdata}B shows the data for one
repetition. The data shows clear clustering around the two regions of
activation, while still allowing for spurious activations in the rest
of the image. This corresponds with the behavior of standard
meta-analysis data (see, e.g., Figure \ref{figdata}).

%
%f4 #&#
\begin{figure}[b]

\includegraphics{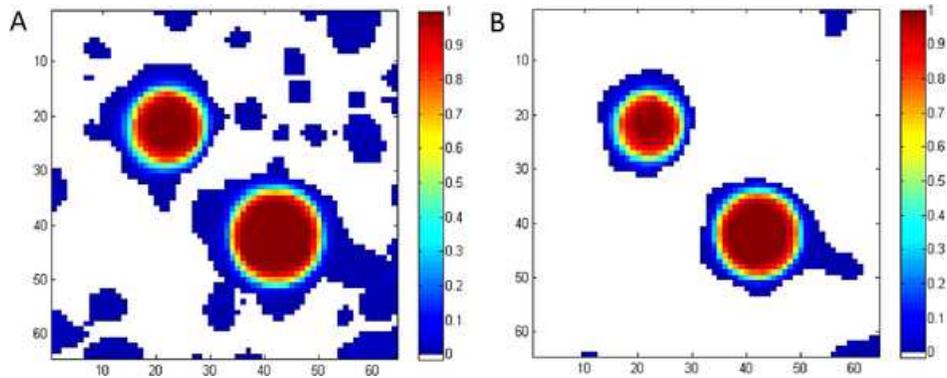}

\caption{Proportion of times each voxel
was deemed significant at the $5\%$ level \textup{(A)} and the $1\%$ level
\textup{(B)} using the ALE method.}
\label{figsimresults}
\end{figure}

\begin{figure}[t!]

\includegraphics{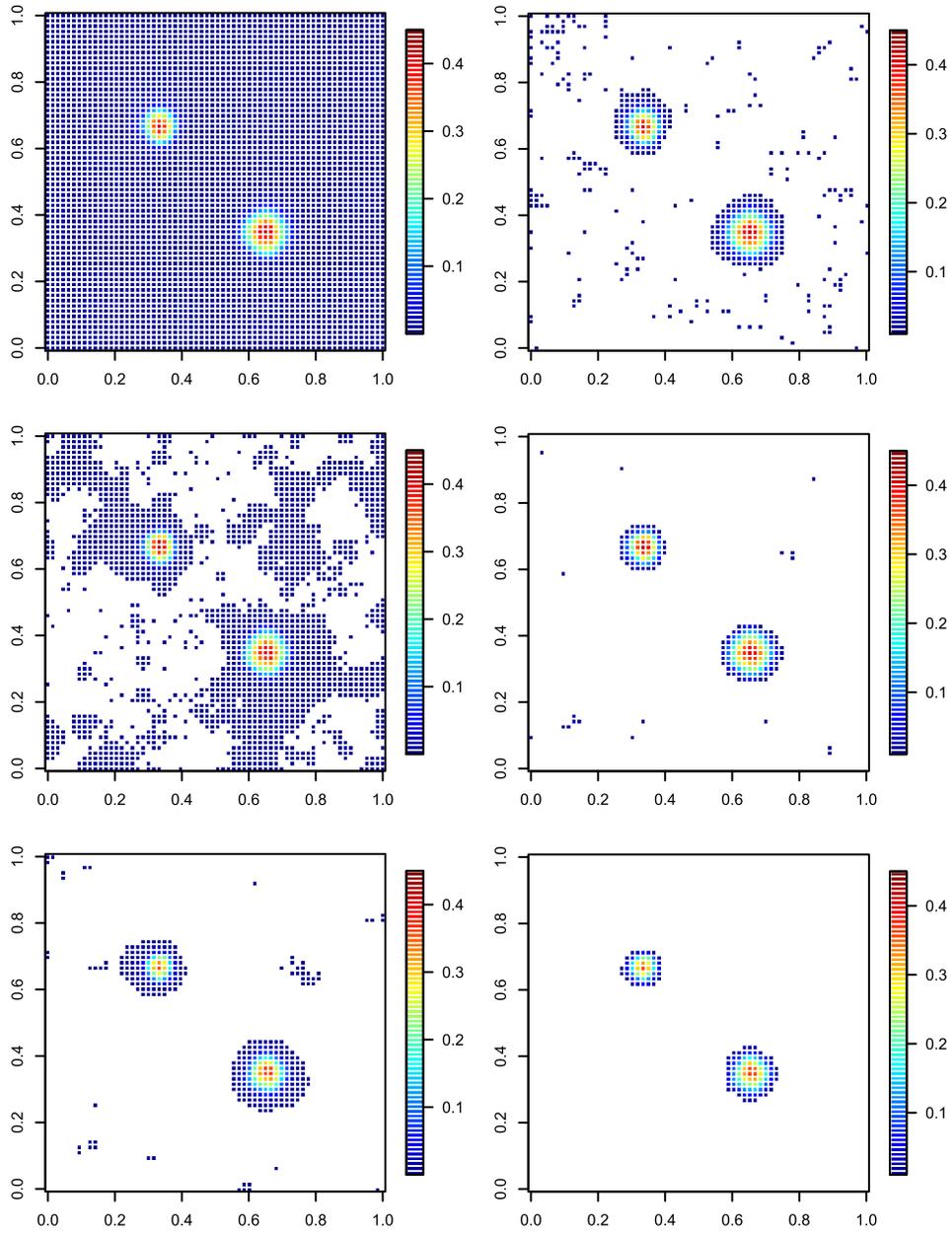}

\caption{Average probability of activation in each voxel obtained
using the adaptive GMRF method combined with the robustification procedure
under different prior miscoding probabilities: $r=0$ (top row),
$r=0.01$ (middle row) and $r=0.05$ (bottom row); The maps in the left
column are not thresholded, while those in the right column are
thresholded at 0.01.}\label{figsim2gmrf}
\end{figure}

Each of the $100$ repetitions were analyzed using the kernel-based ALE
method as well as our Bayesian nonparametric binary regression model.
In the former, a kernel with bandwidth $10\mbox{ mm}$ full width at half
maximum (FWHM) was used, as this is the standard in the field. A Monte
Carlo procedure was used to determine the appropriate threshold to test
the null hypothesis that the reported peak coordinates are uniformly
distributed throughout the grey matter. A permutation distribution is
computed by repeatedly generating peaks at random locations and
performing the smoothing operation to obtain a series of statistical
maps under the null hypothesis that can be used to determine which
voxels had $p$-values below $\alpha$, where $\alpha$ was set to $0.05$
and $0.01$.
Regarding our Bayesian method, the robustification procedure described
in Section \ref{subsecrobustification} is implemented since we use the
background probability of 0.01 to produce the false positives. To see
how sensitive the results are to the use of robustification, we fit the
model with prior miscoding probability $r=0$ (no robustification),
$r=0.01$ and $r=0.05$. Figures \ref{figsimresults}A and B show the
proportion of times each voxel
was deemed significant at the $5\%$ level and the $1\%$ level,
respectively, in the $100$
repetitions, when the ALE method was used. It is clear that the
kernel smoother does a very good job of finding true positives, but
tends to have a large number of false positives in the area
immediately surrounding the activated regions. Figure
\ref{figsim2gmrf} shows the average probability of activation in
each voxel obtained using our method. The maps in the left column are
not thresholded, while those in the right column are thresholded at
0.01. Apparently, our estimates
are closer to the simulated probability map and produce much fewer
false positives than the kernel estimates. Furthermore, our method
yields fewer false positives as the value of $r$, the prior miscoding
probability, increases, that is, the fit becomes more robust. The
spatial extent of the activation region, however, is not significantly
shrunk, making a strong case for the use of adaptive smoothing.
%
%f5 #&#

%
%f6 #&#
\begin{figure}

\includegraphics{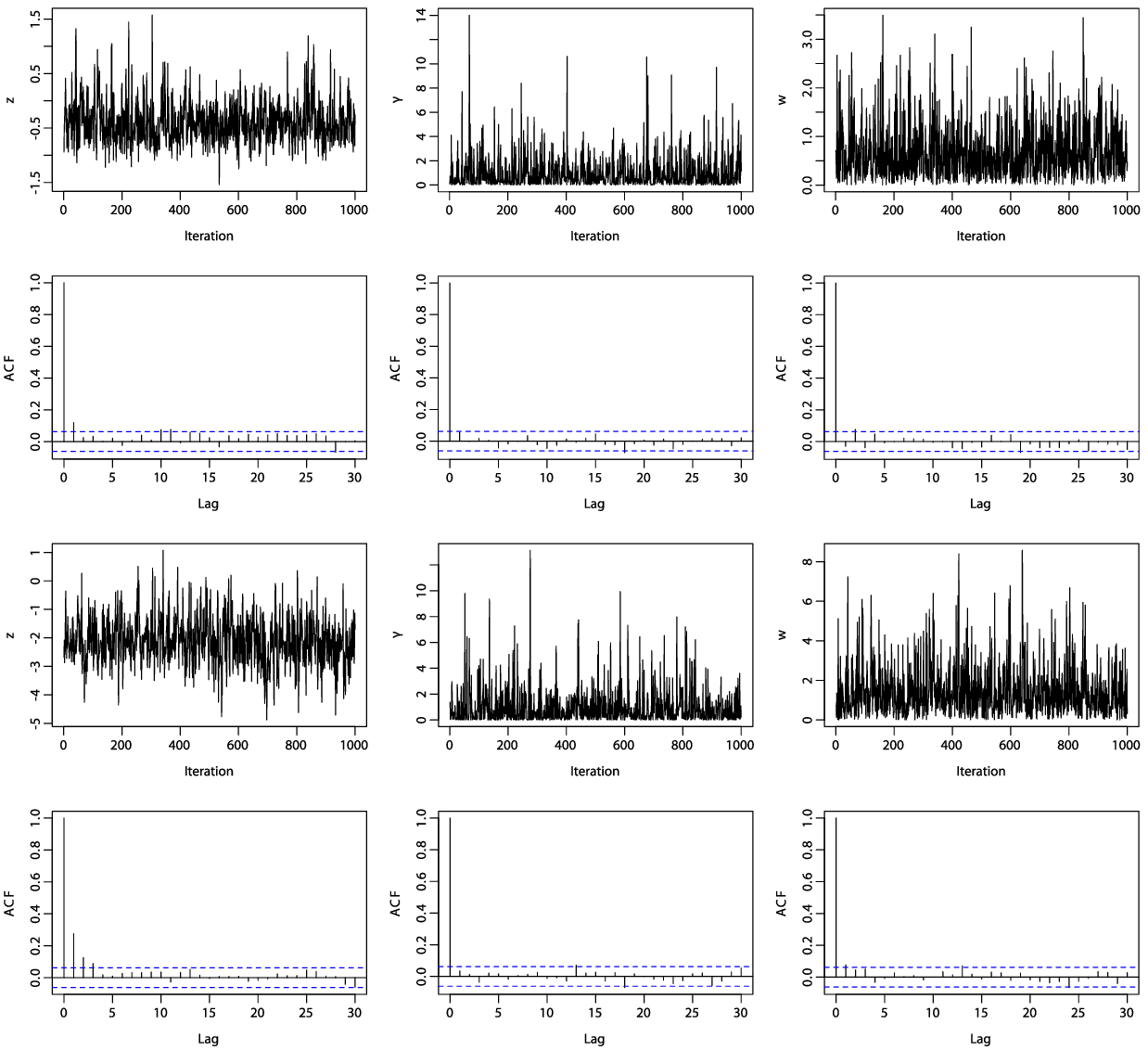}

\caption{Assessment of MCMC convergence for Simulation I. The top
(bottom) two rows contain the typical trace plots and autocorrelation
functions of the samples of variables $z$, $\gamma$ and $w$ from
fitting a probit (logistic) model.}\label{figmcmcdiag}
\end{figure}

%s3.3 #&#
\subsection{Computational performance and MCMC diagnostics}\label{sec3.3}
Thanks to the sparse structure of the adaptive GMRF prior used, the
proposed models provide fast MCMC computation for nonparametric binary
regression. To complete 5000 iterations on a 3.06 GHz Intel iMac
desktop with 4GB memory, it took the probit model 9.23, 46.06 and
11.17 seconds at sample size $n=30\times30$, $60\times60$ and
$90\times90$, respectively, for estimating the bimodal function in
Simulation I. The logistic model is a little slower, taking 11.89,
55.83 and 138.69 seconds to finish the same amount of
computations. The computing times of both models increase with sample
sizes at order $n$, roughly. The programs were written in the FORTRAN
language, making use of the LAPACK and BLAS packages.

It is well known that the GMRF $\bfz$ are strongly dependent on each
other as well as on the auxiliary variable $\bfw$ [see, e.g.,
\citet
{GMRFbook}; \citet{holmheld06}]. Those posterior correlations are
likely to cause slow mixing in the Markov chain. To combat this issue,
we sampled $\bfz$ as a block and employed the joint updating tricks as
used in \citet{holmheld06} (see the \hyperref[appendix]{Appendix} for
details). Since the computation is fast, we also suggest running a
relatively large number of MCMC iterations and applying a thinning
factor of $\ell$ by collecting samples after every $\ell$
iterations. In Simulation I, for instance, we found that it is
sufficient to run 15,000 MCMC iterations (5000 burn-in and 10,000
sampling) with a thinning factor of 10 to obtain reliable
estimates. Figure \ref{figmcmcdiag} shows typical trace plots and
autocorrelation functions of the samples of different variables for
Simulation I. As we can see, the mixing of the chain is satisfactory
for both probit and logistic models.

%
%f7 #&#
\begin{figure}

\includegraphics{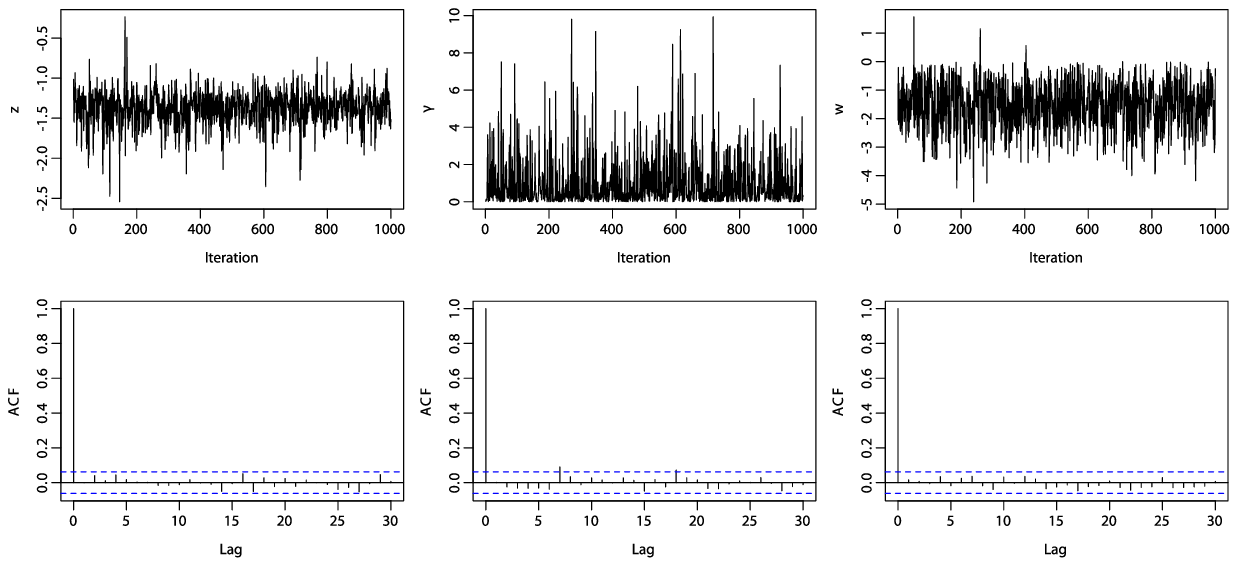}

\caption{Assessment of MCMC convergence for the data analysis. The top
row contains the typical trace plots of the samples of variables $z$,
$\gamma$ and $w$; the bottom row contains the corresponding
autocorrelation functions.}\label{figmcmcmeta}
\end{figure}

%s4 #&#
\section{Data analysis}\label{sec4}\label{sectanalysis}

We describe here the results of our meta-analysis of the fMRI data. As
mentioned before, the data consists of the coordinates of
$2478$ peaks representing the locations of voxel activations,
collected from 162 neuroimaging studies.
The raw data consists of a three-dimensional image with
dimensions $91 \times109 \times91$ whose elements took the value $1$
if an activation\vadjust{\goodbreak} had been reported at that voxel and $0$ otherwise.
Figure \ref{figdata} shows the raw data for a representative slice of the
brain with fixed $x$, $y$ and~$z$ directions, respectively.

The binary nature of the meta-analysis data makes it an ideal
candidate for our Bayesian nonparametric binary regression method.
As our method is currently only implemented in two dimensions, we fit
our method slice-wise across the brain for each orientation (i.e., for
the fixed $x$, $y$ and $z$ direction). Prior to performing our method on
a slice, we applied smoothing in the fixed direction by including all
activations located within $10$~mm of the slice of interest.

In our simulation studies (Section \ref{sectsimstudy}), we found that
the binary regression model is not
sensitive to the choice of link function. We therefore fit a probit
model to the data for computational efficiency. To make our
estimation robust against false\vadjust{\goodbreak} positives, we
incorporated the robustification procedure (Section~\ref{sec4}) in the model
with prior miscoding probability $r = 0.01$ for every voxel. Due to the
high dimension of the data, the MCMC
was run for 60,000 iterations with 10,000 burn-in and a thinning factor
of 50 iterations, resulting in posterior samples of size 1000. The
Markov chains mix well as shown in Figure \ref{figmcmcmeta}.

%
%f8 #&#
\begin{figure}[b]

\includegraphics{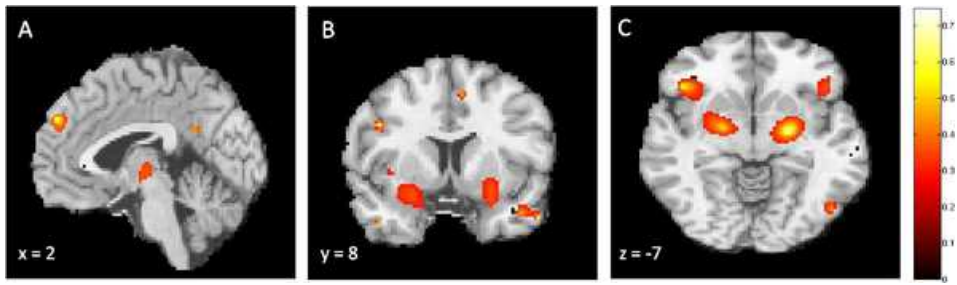}

\caption{Thresholded posterior probability maps are shown for the
sagittal, coronal and axial slice of the brain depicted in Figure
\protect\ref
{figdata}. Regions with posterior probability of observing a peak
activation higher than $0.30$ are color-coded.}
\label{figresults}
\end{figure}

%
%f9 #&#
\begin{figure}

\includegraphics{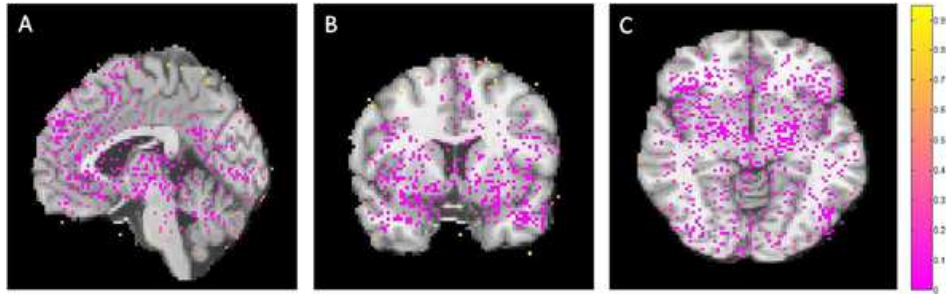}

\caption{ Miscoding probabilities are shown for the sagittal, coronal
and axial slice of the brain depicted in Figure \protect\ref
{figdata}. Points
with posterior miscoding probability higher than $0.10$ are color-coded.}
\label{figmiscoding}
\end{figure}

Once the Bayesian binary regression model was fit, posterior
probability maps were obtained indicating the probability of being a
location of peak activation across the brain. Regions with probability
values higher than $0.3$ were color-coded and superimposed onto an
anatomical reference image. The relatively low threshold is indicative
of the dispersion of foci locations in the data. Figure \ref
{figresults} shows results for the three slices described above. Key
regions of activation observed in the figure include the thalamus (8A),
amygdala (8B)\vadjust{\goodbreak} and the ventral striatum (8C). These regions are known to
be associated with emotion, and were also indicated as active when
using kernel-based methods [see \citet{Kober2008}]. It should be noted
we obtain the same regions of activation as \citet{Kober2008},
but with
significantly smaller spatial extent. This is consistent with our
simulation study, which shows how the kernel-based methods tend to
overestimate the extent of activation. Finally, Figure \ref
{figmiscoding} shows the posterior miscoding probabilities (thresholded
at 0.10) for the same three slices. High miscoding probabilities
indicate points that were deemed to be spurious activations and
therefore given lower weights when calculating the posterior activation
probabilities. Based on their locations, it appears that our method is
providing an effective means of downweighting false activations.

To see if the adaptive smoothing is preferred to the ordinary smoothing
in this neuroimaging example, we conducted a test on $H_{0}\dvtx \gamma_{jk}
= 1$ using the deviance information criterion (DIC) introduced by
\citet
{spiegdic02}. More specifically, we first fitted to our imaging data
the proposed adaptive GMRF model and a (nonadaptive) Bayesian
thin-plate spline model (by fixing all $\gamma_{jk}$ to be 1), and
saved the MCMC posterior samples of both models. Then, we define the
deviance as $D(\bfphi) = -2\log(p(\bfy|\bfphi))$, where $p(\bfy
|\bfphi
)$ is the likelihood function and $\bfphi$ are unknown parameters of
the model. The DIC score is finally estimated using $\mathit{DIC} = 2\bar{D} -
D(\bar{\bfphi})$, where $\bar{D}$ is calculated as the average of
$D(\bfphi)$ over the samples of $\bfphi$, and $D(\bar{\bfphi})$ as the
value of $D$ evaluated at the average of the samples of $\bfphi$. The
model with smaller DIC should be in favor. Table \ref{tabdic} shows the
DIC scores of the two models for the fixed $x$, $y$ and $z$
orientations, where the adaptive model is preferred in every scenario.

%
%t1 #&#
\begin{table}
\tabcolsep=0pt
\tablewidth=217pt
\caption{DIC scores of both adaptive and nonadaptive models for the
fixed $x$, $y$ and $z$ orientation.}\label{tabdic}
\begin{tabular*}{217pt}{@{\extracolsep{\fill}}lccc@{}}
\hline
\textbf{Orientation}&\textbf{x}&\textbf{y}&\textbf{z}\\
\hline
Adaptive & \hphantom{0\mbox{,}}9918.372 & 8216.255 & 9917.209\\
Nonadaptive &10\mbox{,}090.460 & 9512.016 & 9947.377\\
\hline
\end{tabular*}
\end{table}

%s5 #&#
\section{Discussion}\label{sec5} \label{sectsummary}

We developed a fully Bayesian method for nonparametric binary
regression and, together with a robustification procedure,
applied it to meta-analysis in fMRI studies. Our
analysis identified activated regions\vadjust{\goodbreak} of the brain that are known to
be associated with emotion. While similar regions were also identified
in other meta-analyses such as \citet{Kober2008} that use kernel-based
methods, our method has several advantages over such approaches as
follows. The adaptive GMRF used in our model better matches
the natural spatial resolution of the data across the brain compared
to using an arbitrary chosen fixed kernel size. This allows us to avoid
the problem of overestimating regions of activation apparent in
kernel-based methods. The Bayesian nature of
our method allows for the construction of posterior probability maps
indicating the probability of observing a peak activation in response
to the paradigm across the brain. This is in contrast to kernel methods
which simply state that more peaks lie near the voxel than expected by
chance. It should be noted that recently a Bayesian spatial
hierarchical model using a marked independent cluster process
[\citet
{KangJASA2011}] was introduced for dealing with neuroimaging
meta-analysis. In future work we will look at comparing this method
with the nonparametric binary regression approach suggested in this
paper. Finally, our procedure provides estimates of miscoding
probabilities which can help to identify regions that may have been
incorrectly tagged as being activated. This is another feature not
provided by kernel-based methods.

It is important to note that in this work the model setup assumes that
the input data is two dimensional. Such 2D smoothing serves a useful
purpose, as fMRI data are often analyzed either slice-wise or using
cortical surface-based techniques [\citet{Dale1999}; \citet
{Fischl1999}]. In
reality, however, fMRI data are three dimensional in space. Therefore,
it may ultimately be more appropriate to smooth the three spatial
dimensions directly. We are actually working on such an extension of
our current approach. The main computational constraint stems from
inverting a large precision matrix, which is of
$91\times109\times91=902{,}629$ dimensions in our neuroimaging
example. We thus need a practical 3D GMRF, but the construction is
nontrivial. One possible solution is to obtain a highly sparse
precision matrix by discretizing a 3D Laplacian operator with proper
boundary conditions as we did in the 2D case. To achieve more
computational efficiency, we may use a novel Bayesian inference tool
similar to that introduced in \citet{rueinla09} rather than MCMC.

As shown in the simulation studies, the results obtained by our method
are somewhat sensitive to the prior miscoding probability~$r$\vadjust{\goodbreak} in the
robustification procedure. A large $r$ may underestimate the activation
clusters, while a small~$r$ tends to allow more false positives. The
choice of~$r$ is often subjective. One may use information from, say,
previous studies, to find an appropriate~$r$ in order to balance this
trade-off. If no prior information is available, \citet{choughoroy07}
proposed letting $r$ be a small number between 0.01 and 0.1. In
practice, we suggest experimenting with several $r$ values and choosing
the one that gives the best results.

%Our Bayesian nonparametric binary regression method can be used in
%other settings. By using locally adaptive Gaussian Markov random
%fields we
%allow for locally adaptive smoothing of the probability function with
%respect to the covariates. Our simulation study shows that the method
%is computationally efficient and performs very well compared with
%other available methods for nonparametric binary regression.
%
%Extensions?

%Besides being locally adaptive, our procedure also provides estimates
%of miscoding probabilities which can help to identify regions that may
%have been incorrectly tagged as being activated. Furthermore,
%posterior probability activation maps are also produced.
%

%apA #&#
\begin{appendix}
%sB #&#
\section*{Appendix: MCMC algorithms for posterior inference}\label{secB}\label{appendix}

%sB.1 #&#
\subsection{Probit link}\label{secB.1}\label{subsecprobit}

Let $\bfy= (y_1,\ldots,y_n)^T$ be the random vector of binary
observations measured and $\bfx=(x_1,\ldots,x_n)^T$ the corresponding
covariate values, where each $x_i$ has one or two component variables.
Let $\bfw=(w_1,\ldots,\allowbreak w_n)^T$ be some unobservable latent variable.
Following \citet{holmheld06}, the probit model can be written as
%
%e9 #&#
%e10 #&#
\begin{eqnarray}\label{modprob}
 y_i&=&
\cases{\displaystyle
1, &  \quad  if  $w_i>0$,\cr\displaystyle
0, & \quad   if  $w_i\le0$,}
\nonumber
\\[-8pt]
\\[-8pt]
 w_i &=& z_i + \vareps_i, \qquad \vareps_i\stackrel{\mathrm{i.i.d.}}{\sim}\mathrm{N}(0,1),
\nonumber
\end{eqnarray}
where $\bfz=(z_1,\ldots,z_n)^T$ is the adaptive GMRF described in
Section \ref{subsecspatialadaptive}.
%[\bfz|\delta^2,\bfgamma]\propto|\delta^{-2}\bfA_\gamma|^{1/2}_+\exp
% (-\fr{1}{2\delta^2}\bfz^T\bfA_\gamma\bfz),
%where $\bfA_\gamma=\bfB_m'\mbox{diag}(\bfgamma)\bfB_m$ for $m=1,2$.
Since~$y_i$ are now deterministic conditional on the sign of the $w_i$,
we have $P(y_i=0| z_i) =P(w_i\le0| z_i)= \Phi(-z_i)$, where
$\Phi
$ is the standard Gaussian c.d.f.

As mentioned earlier, the half-$t$ prior on $\delta$ can be written as
$\delta\stackrel{\mathcal{D}}{=}|\xi|\theta$, where $\xi\sim
\mathrm
{N}(0,1)$ and $\theta^2\sim\operatorname{IG}(\rho/2,\rho S^2/2)$. A redundant
multiplicative reparameterization can be applied to model (\ref{modprob}):
\begin{eqnarray*}
 y_i&=&
\cases{\displaystyle
1, & \quad   if  $w_i>0$,\cr\displaystyle
0, & \quad   if  $w_i\le0$,}\\
 w_i &=& \xi\eta_i + \vareps_i, \qquad \vareps_i\stackrel{\mathrm{i.i.d.}}{\sim
}\mathrm{N}(0,1),
\end{eqnarray*}
where $\bfeta=(\eta_1,\ldots,\eta_n)^T$ has a GMRF prior density
\begin{eqnarray*}
[\bfeta|\theta^2,\bfgamma]\propto|\theta^{-2}\bfA_\gamma
|^{1/2}_+\exp
\biggl(-\frac{1}{2\theta^2}\bfeta^T\bfA_\gamma\bfeta\biggr),
\end{eqnarray*}
with $\bfA_\gamma=\bfB_m'\operatorname{diag}(\bfgamma)\bfB_m$ for
$m=1,2$. This
expanded model form allows conditionally conjugate
prior distributions for both $\xi$ and $\theta$, and these parameters
are independent in the conditional posterior distribution [\citet
{Gelmprio2006}; \citet{gelmanjcgs08}]. Letting $d$ be the
dimension of the null
space of $\bfA_\gamma$, the full conditional distributions are listed below:
\begin{itemize}
\item
$(\bfeta|\theta^2,\xi,\bfgamma,\bfw)\sim\mathrm{N}_n(\bfmu
_\eta
,\bfSigma_\eta)$, where $\bfmu_\eta=\xi\bfSigma_\eta\bfw$ and
$\bfSigma
_\eta=(\xi^2\bfI_n+\break\bfA_\gamma/ \theta^2)^{-1}$;
\item
$(\xi|\bfeta,\bfw)\sim\mathrm{N} (\mu_\xi,\sigma_\xi^2 )$,
where $\mu_\xi=\sigma_\xi^2\bfeta'\bfw$ and $\sigma_\xi
^2=(1+\bfeta
'\bfeta)^{-1}$;\vspace*{2pt}
\item
$(w_i|\xi,\bfeta,\bfy)\sim
\cases{\displaystyle
 N (\xi\eta_i,1)I(w_i>0), & \quad   if  $y_i=1$,\cr\displaystyle
 N (\xi\eta_i,1)I(w_i\le0), & \quad   if  $y_i=0$;}
$\vspace*{2pt}
\item
$(\gamma_j|\theta^2,\bfeta)\sim\operatorname{IG} (\frac{\nu
+1}{2},\frac
{1}{2\theta^2}\tilde{\eta}_j^2+\frac{1}{2} )$, where $\tilde
{\bfeta
}=\bfB_m\bfeta$ $(m=1,2)$;
\item
$(\theta^2|\bfeta,\bfgamma)\sim\operatorname{IG} (\frac{n-d}{2}+\frac
{\rho
}{2},\frac{1}{2}\bfeta'\bfA_\gamma\bfeta+\frac{\rho S^2}{2} )$.
\end{itemize}
Note that $\bfSigma_\eta$ is a banded matrix and we can thus use
the banded Cholesky decomposition to simulate $\bfeta$ with the cost of
$\mathcal{O}(n)$. The quantities
$w_i$ have independent truncated normal distributions and are also
straightforward to sample from.

%sB.2 #&#
\subsection{Logit link}\label{secB.2}\label{subseclogit}
Again, we use data augmentation and overparameterization to write the
logistic regression model as
%
%e11 #&#
%e12 #&#
\begin{eqnarray}\label{modlogisticexp}
 y_i&=&
\cases{
1, &   \quad {if} $w_i>0$,\cr
0, &  \quad {if} $w_i\le0$,
}
\nonumber\\
 w_i &=& \xi\eta_i + \vareps_i, \qquad \vareps_i\sim\mathrm
{N}(0,\lambda
_i),\\
%&&\nabla^m z_j\stackrel{\mathrm{i.i.d.}}{\sim}\mathrm{N}(0,\theta^2\gamma_j^2),\quad
%j=m+1,\ldots,n,\\
 \lambda_i &=& (2\kappa_i)^2, \qquad \kappa_i\sim\mathrm{KS},\nonumber
%&&\xi\sim\mathrm{N}(0,1),\quad\gamma^2_j\sim\mbox{IG} (\fr{\nu}{2},
\end{eqnarray}
where KS denotes the Kolmogorov--Smirnov distribution
[e.g., \citet{Devr1986}]. In this case, $\vareps_i$ has the form of
a scale mixture of normals with a marginal logistic distribution.
%so that the marginal likelihood for models (\ref{modlogisticexp}) and
%(??) are equivalent.

To improve mixing of the Markov chains, we update $\{\bfw,\boldsymbol
\lambda\}$
jointly given $\{\xi,\bfeta\}$,
\begin{eqnarray*}
[\bfw,\boldsymbol\lambda|\xi,\bfeta,\bfy]=[\bfw|\xi
,\bfeta,\bfy
][\boldsymbol\lambda
|\bfw,\xi,\bfeta].
\end{eqnarray*}
Letting $\bfLambda=\operatorname{diag}(\lambda_1,\ldots,\lambda_n)$, the
posterior conditional distributions are as follows:
\begin{itemize}
\item
$(\bfeta|\theta^2,\xi,\bfgamma,\bfw,\boldsymbol\lambda)\sim
\mathrm
{N}_n(\bfmu
_\eta,\bfSigma_\eta)$, where $\bfmu_\eta=\xi\bfSigma_\eta
\bfLambda\bfw$
and $\bfSigma_z=(\xi^2\bfLambda+\break\bfA_\gamma/\theta^2)^{-1}$;
\item
$(\xi|\bfeta,\bfw,\boldsymbol\lambda)\sim\mathrm{N} (\mu_\xi
,\sigma
_\xi
^2 )$, where $\mu_\xi=\sigma_\xi^2\bfeta'\bfLambda\bfw$ and
$\sigma
_\xi^2=(1+\bfeta'\bfLambda\bfeta)^{-1}$;\vspace*{2pt}
\item
$(w_i|\xi,\bfeta,\bfy)\sim
\cases{\displaystyle
 \operatorname{Logistic} (\xi\eta_i,1)I(w_i>0), & \quad   if  $y_i=1$,\cr\displaystyle
 \operatorname{Logistic} (\xi\eta_i,1)I(w_i\le0), & \quad   if  $y_i=0$;}
$\vspace*{1pt}
\item
$[\lambda_i| w_i,\xi,\eta_i]\propto\lambda_i^{-1}\exp\{-\frac
{1}{2\lambda_i}(w_i-\xi\eta_i)^2 \}\operatorname{KS} (\frac{\sqrt{\lambda
_i}}{2} )$;
\item
$(\gamma_j|\theta^2,\bfeta)\sim\operatorname{IG} (\frac{\nu
+1}{2},\frac
{1}{2\theta^2}\tilde{\eta}_j^2+\frac{\nu}{2} )$, where $\tilde
{\bfeta
}=\bfB_m\bfeta$ $(m=1,2)$;
\item
$(\theta^2|\bfeta,\bfgamma)\sim\operatorname{IG} (\frac{n-d}{2}+\frac
{\rho
}{2},\frac{1}{2}\bfeta'\bfA_\gamma\bfeta+\frac{\rho S^2}{2} )$.
\end{itemize}
The $\operatorname{Logistic}(\alpha,\beta)$ denotes the density function of
the logistic distribution with mean $\alpha$ and scale parameter
$\beta$ [\citet{Devr1986}, page 39]. Sampling from the truncated logistic
distribution can be done efficiently by the inversion method. Although
it is not a standard task, sampling $\lambda_i$ is simple using a
rejection method as outlined in \citet{holmheld06}.

%sB.3 #&#
\subsection{Other scale mixtures of normal links}\label{secB.3}\label{subsecsmn}
The auxiliary variable sampling scheme described above can easily be
generalized to work for any link function $H$ that can be represented
as scale mixtures of normal cdfs, and, hence,
\begin{eqnarray*}
H(t) = \int_0^\infty\Phi\biggl(\frac{t}{\sqrt{v}} \biggr)\,dG(v),
\end{eqnarray*}
where $v$ follows some continuous or discrete distribution $G$ on
$(0,\infty)$. A wide class of continuous, unimodal and symmetric
distributions on the real line may be constructed as scale mixtures of
normals. Many examples, such as discrete mixtures or contaminated
normals, the Student $t$ family, logistic, Laplace or
double-exponential, and the stable family, are well known; see, for
example, \citet{andmal74}.

Similarly, we introduce two sets of latent variables $\bfw=(w_1,\ldots
,w_n)^T$ and $\mathbf v=(v_1,\ldots,v_n)^T$ such that $(w_i|\bfz
,\mathbf v
)\sim\mathrm{N}(z_i,v_i)$, $v_i\stackrel{\mathrm{i.i.d.}}{\sim}G$, and $y_i =
I(w_i>0)$. Then, conditional on $\bfz$, the $y_i$'s are independent
Bernoulli random variables with success probability $H(z_i)$. Suppose
$G$ has a Lebesgue density or probability mass function $g$. Let $z_i =
\xi\eta_i$ and $\bfV= \operatorname{diag}(v_1,\ldots,v_n)$. Then, the
posterior conditional distributions are
as follows:
\begin{itemize}
\item
$(\bfeta|\theta^2,\xi,\bfgamma,\bfw,\mathbf v)\sim\mathrm
{N}_n(\bfmu_\eta
,\bfSigma_\eta)$, where $\bfmu_\eta=\xi\bfV\bfSigma_\eta\bfw$ and
$\bfSigma_\eta=(\xi^2\bfV+\bfA_\gamma/\theta^2)^{-1}$;
\item
$(\xi|\bfeta,\bfw,\mathbf v)\sim\mathrm{N} (\mu_\xi,\sigma
_\xi^2
)$, where $\mu_\xi=\sigma_\xi^2\bfeta'\bfV\bfw$ and $\sigma_\xi
^2=(1+\bfeta'\bfV\bfeta)^{-1}$;\vspace*{2pt}
\item
$(w_i|\xi,\bfeta,\mathbf v,\bfy)\sim
\cases{\displaystyle
 \mathrm{N} (\xi\eta_i,v_i)I(w_i>0), & \quad   if  $y_i=1$,\cr\displaystyle
 \mathrm{N} (\xi\eta_i,v_i)I(w_i\le0), & \quad   if  $y_i=0$;}
$\vspace*{2pt}
\item
$[v_i|\xi, w_i,\eta_i]\propto v_i^{-1/2}\exp\{-\frac
{1}{2v_i}(w_i-\xi\eta_i)^2 \}g(v_i)$;
\item
$(\gamma_j|\theta^2,\bfeta)\sim\operatorname{IG} (\frac{\nu
+1}{2},\frac
{1}{2\theta^2}\tilde{\eta}_j^2+\frac{\nu}{2} )$, where $\tilde
{\bfeta
}=\bfB_m\bfeta$ $(m=1,2)$;
\item
$(\theta^2|\bfeta,\bfgamma)\sim\operatorname{IG} (\frac{n-d}{2}+\frac
{\rho
}{2},\frac{1}{2}\bfeta'\bfA_\gamma\bfeta+\frac{\rho S^2}{2} )$.
\end{itemize}
Thus, a Gibbs sampler can be used to sample joint posterior
distributions. The only difficult part is sampling $\theta_i$. For a
Student $t$ link, the mixing distribution $G$ is an inverse gamma
distribution, as is the full conditional of each~$v_i$. For the Laplace
link, the $G$ is an exponential distribution and the~$v_i^{-1}$ follows
an inverse Gaussian conditional distribution. Therefore, one can
directly sample $v_i$'s for those two links. If $[v_i|\xi,
w_i,\eta
_i]$ does not correspond to any regular density, the samples may be
drawn via acceptance-rejection sampling.
\end{appendix}

\section*{Acknowledgment}
The authors thank Tor Wager for the meta-analysis data.
We would like to thank the anonymous referees and Associate Editor for
valuable comments that led to considerable improvement upon the original
version of this paper. We also thank Tor Wager for the meta-analysis data.

%suskaldyti doi

% imsref loaded by smiklovaite, 2012-01-11 07:28:51
% imsref loaded by smiklovaite, 2012-01-11 07:34:29
%

\printaddresses


\begin{thebibliography}{46}
% BibTex style file: ims.bst, 2011-05-30
% Default style options (sort=0,type=number).
% Used options (sort=1,type=nameyear).

%b1 #&#
\bibitem[\protect\citeauthoryear{Albert and Chib}{1993}]{albechib93}
%
\begin{barticle}[mr]
\bauthor{\bsnm{Albert},~\bfnm{James~H.}\binits{J.~H.}} \AND
\bauthor{\bsnm{Chib},~\bfnm{Siddhartha}\binits{S.}}
(\byear{1993}).
\btitle{Bayesian analysis of binary and polychotomous response data}.
\bjournal{J. Amer. Statist. Assoc.}
\bvolume{88}
\bpages{669--679}.
\bid{issn={0162-1459}, mr={1224394}}
\bptok{imsref}%
\end{barticle}
%
\endbibitem

%b2 #&#
\bibitem[\protect\citeauthoryear{Andrews and Mallows}{1974}]{andmal74}
%
\begin{barticle}[mr]
\bauthor{\bsnm{Andrews},~\bfnm{D.~F.}\binits{D.~F.}} \AND
\bauthor{\bsnm{Mallows},~\bfnm{C.~L.}\binits{C.~L.}}
(\byear{1974}).
\btitle{Scale mixtures of normal distributions}.
\bjournal{J.~Roy. Statist. Soc. Ser. B}
\bvolume{36}
\bpages{99--102}.
\bid{issn={0035-9246}, mr={0359122}}
\bptok{imsref}%
\end{barticle}
%
\endbibitem

%b3 #&#
\bibitem[\protect\citeauthoryear{Brezger, Fahrmeir and
Hennerfeind}{2007}]{BrezgerFahmhen2007}
%
\begin{barticle}[mr]
\bauthor{\bsnm{Brezger},~\bfnm{A.}\binits{A.}},
\bauthor{\bsnm{Fahrmeir},~\bfnm{L.}\binits{L.}} \AND
\bauthor{\bsnm{Hennerfeind},~\bfnm{A.}\binits{A.}}
(\byear{2007}).
\btitle{Adaptive {G}aussian {M}arkov random fields with applications
in human
brain mapping}.
\bjournal{J. Roy. Statist. Soc. Ser. C}
\bvolume{56}
\bpages{327--345}.
\bid{doi={10.1111/j.1467-9876.2007.00580.x}, issn={0035-9254}, mr={2370993}}
\bptok{imsref}%
\end{barticle}
%
\endbibitem

%b4 #&#
\bibitem[\protect\citeauthoryear{Carter and Kohn}{1996}]{CartKohnmark1996}
%
\begin{barticle}[mr]
\bauthor{\bsnm{Carter},~\bfnm{C.~K.}\binits{C.~K.}} \AND
\bauthor{\bsnm{Kohn},~\bfnm{R.}\binits{R.}}
(\byear{1996}).
\btitle{Markov chain {M}onte {C}arlo in conditionally {G}aussian state space
models}.
\bjournal{Biometrika}
\bvolume{83}
\bpages{589--601}.
\bid{doi={10.1093/biomet/83.3.589}, issn={0006-3444}, mr={1423876}}
\bptok{imsref}%
\end{barticle}
%
\endbibitem

%b5 #&#
\bibitem[\protect\citeauthoryear{Carvalho, Polson and
Scott}{2010}]{CarvPolsScothors2010}
%
\begin{barticle}[mr]
\bauthor{\bsnm{Carvalho},~\bfnm{Carlos~M.}\binits{C.~M.}},
\bauthor{\bsnm{Polson},~\bfnm{Nicholas~G.}\binits{N.~G.}} \AND
\bauthor{\bsnm{Scott},~\bfnm{James~G.}\binits{J.~G.}}
(\byear{2010}).
\btitle{The horseshoe estimator for sparse signals}.
\bjournal{Biometrika}
\bvolume{97}
\bpages{465--480}.
\bid{doi={10.1093/biomet/asq017}, issn={0006-3444}, mr={2650751}}
\bptok{imsref}%
\end{barticle}
%
\endbibitem

%b6 #&#
\bibitem[\protect\citeauthoryear{Choudhuri, Ghosal and
Roy}{2007}]{choughoroy07}
%
\begin{barticle}[mr]
\bauthor{\bsnm{Choudhuri},~\bfnm{Nidhan}\binits{N.}},
\bauthor{\bsnm{Ghosal},~\bfnm{Subhashis}\binits{S.}} \AND
\bauthor{\bsnm{Roy},~\bfnm{Anindya}\binits{A.}}
(\byear{2007}).
\btitle{Nonparametric binary regression using a {G}aussian process prior}.
\bjournal{Stat. Methodol.}
\bvolume{4}
\bpages{227--243}.
\bid{doi={10.1016/j.stamet.2006.07.003}, issn={1572-3127}, mr={2368147}}
\bptok{imsref}%
\end{barticle}
%
\endbibitem

%b7 #&#
\bibitem[\protect\citeauthoryear{Crainiceanu et~al.}{2007}]{CraiRupp07}
%
\begin{barticle}[author]
\bauthor{\bsnm{Crainiceanu},~\bfnm{C.~M.}\binits{C.~M.}},
\bauthor{\bsnm{Ruppert},~\bfnm{D.}\binits{D.}},
\bauthor{\bsnm{Carroll},~\bfnm{R.~J.}\binits{R.~J.}},
\bauthor{\bsnm{Adarsh},~\bfnm{J.}\binits{J.}} \AND
\bauthor{\bsnm{Goodner},~\bfnm{B.}\binits{B.}}
(\byear{2007}).
\btitle{Spatially adaptive penalized splines with heteroscedastic errors}.
\bjournal{J. Comput. Graph. Statist.}
\bvolume{16}
\bpages{265--288}.
\bptok{imsref}%
\end{barticle}
%
\endbibitem

%b8 #&#
\bibitem[\protect\citeauthoryear{Dale, Fischl and Sereno}{1999}]{Dale1999}
%
\begin{barticle}[pbm]
\bauthor{\bsnm{Dale},~\bfnm{A.~M.}\binits{A.~M.}},
\bauthor{\bsnm{Fischl},~\bfnm{B.}\binits{B.}} \AND
\bauthor{\bsnm{Sereno},~\bfnm{M.~I.}\binits{M.~I.}}
(\byear{1999}).
\btitle{Cortical surface-based analysis. I. Segmentation and surface
reconstruction}.
\bjournal{Neuroimage}
\bvolume{9}
\bpages{179--194}.
\bid{doi={10.1006/nimg.1998.0395}, issn={1053-8119},
pii={S1053-8119(98)90395-0}, pmid={9931268}}
\bptok{imsref}%
\end{barticle}
%
\endbibitem

%b9 #&#
\bibitem[\protect\citeauthoryear{Devroye}{1986}]{Devr1986}
%
\begin{bbook}[author]
\bauthor{\bsnm{Devroye},~\bfnm{Luc}\binits{L.}}
(\byear{1986}).
\btitle{Non-Uniform Random Variate Generation}.
\bpublisher{Springer}, \baddress{New York}.
\bptok{imsref}%
\end{bbook}
%
\endbibitem

%b10 #&#
\bibitem[\protect\citeauthoryear{Dey, Ghosh and Mallick}{2000}]{deyghomal99}
%
\begin{bbook}[mr]
\beditor{\bsnm{Dey},~\bfnm{Dipak~K.}\binits{D.~K.}},
\beditor{\bsnm{Ghosh},~\bfnm{Sujit~K.}\binits{S.~K.}} \AND
\beditor{\bsnm{Mallick},~\bfnm{Bani~K.}\binits{B.~K.}}, eds.
(\byear{2000}).
\btitle{Generalized Linear Models: A Bayesian Perspective}.
\bseries{Biostatistics}
\bvolume{5}.
\bpublisher{Dekker}, \baddress{New York}.
\bid{mr={1893779}}
\bptok{imsref}%
\end{bbook}
%
\endbibitem

%b11 #&#
\bibitem[\protect\citeauthoryear{Fischl, Sereno and Dale}{1999}]{Fischl1999}
%
\begin{barticle}[pbm]
\bauthor{\bsnm{Fischl},~\bfnm{B.}\binits{B.}},
\bauthor{\bsnm{Sereno},~\bfnm{M.~I.}\binits{M.~I.}} \AND
\bauthor{\bsnm{Dale},~\bfnm{A.~M.}\binits{A.~M.}}
(\byear{1999}).
\btitle{Cortical surface-based analysis. II: Inflation, flattening,
and a
surface-based coordinate system}.
\bjournal{Neuroimage}
\bvolume{9}
\bpages{195--207}.
\bid{doi={10.1006/nimg.1998.0396}, issn={1053-8119},
pii={S1053-8119(98)90396-2}, pmid={9931269}}
\bptok{imsref}%
\end{barticle}
%
\endbibitem

%b12 #&#
\bibitem[\protect\citeauthoryear{Gelman}{2006}]{Gelmprio2006}
%
\begin{barticle}[mr]
\bauthor{\bsnm{Gelman},~\bfnm{Andrew}\binits{A.}}
(\byear{2006}).
\btitle{Prior distributions for variance parameters in hierarchical models
(comment on article by {B}rowne and {D}raper)}.
\bjournal{Bayesian Anal.}
\bvolume{1}
\bpages{515--533 (electronic)}.
\bid{mr={2221284}}
\bptnote{check related}%
\bptok{imsref}%
\end{barticle}
%
\endbibitem

%b13 #&#
\bibitem[\protect\citeauthoryear{Gelman et~al.}{2008}]{gelmanjcgs08}
%
\begin{barticle}[mr]
\bauthor{\bsnm{Gelman},~\bfnm{Andrew}\binits{A.}}, \bauthor
{\bparticle{van}
\bsnm{Dyk},~\bfnm{David~A.}\binits{D.~A.}},
\bauthor{\bsnm{Huang},~\bfnm{Zaiying}\binits{Z.}} \AND
\bauthor{\bsnm{Boscardin},~\bfnm{W.~John}\binits{W.~J.}}
(\byear{2008}).
\btitle{Using redundant parameterizations to fit hierarchical models}.
\bjournal{J. Comput. Graph. Statist.}
\bvolume{17}
\bpages{95--122}.
\bid{doi={10.1198/106186008X287337}, issn={1061-8600}, mr={2424797}}
\bptok{imsref}%
\end{barticle}
%
\endbibitem

%b14 #&#
\bibitem[\protect\citeauthoryear{Gu}{1990}]{gu90}
%
\begin{barticle}[mr]
\bauthor{\bsnm{Gu},~\bfnm{Chong}\binits{C.}}
(\byear{1990}).
\btitle{Adaptive spline smoothing in non-{G}aussian regression models}.
\bjournal{J. Amer. Statist. Assoc.}
\bvolume{85}
\bpages{801--807}.
\bid{issn={0162-1459}, mr={1138360}}
\bptok{imsref}%
\end{barticle}
%
\endbibitem

%b15 #&#
\bibitem[\protect\citeauthoryear{Hastie and Tibshirani}{1990}]{Hastie90}
%
\begin{bbook}[mr]
\bauthor{\bsnm{Hastie},~\bfnm{T.~J.}\binits{T.~J.}} \AND
\bauthor{\bsnm{Tibshirani},~\bfnm{R.~J.}\binits{R.~J.}}
(\byear{1990}).
\btitle{Generalized Additive Models}.
\bseries{Monographs on Statistics and Applied Probability}
\bvolume{43}.
\bpublisher{Chapman \& Hall}, \baddress{London}.
\bid{mr={1082147}}
\bptok{imsref}%
\end{bbook}
%
\endbibitem

%b16 #&#
\bibitem[\protect\citeauthoryear{Holmes and Held}{2006}]{holmheld06}
%
\begin{barticle}[mr]
\bauthor{\bsnm{Holmes},~\bfnm{Chris~C.}\binits{C.~C.}} \AND
\bauthor{\bsnm{Held},~\bfnm{Leonhard}\binits{L.}}
(\byear{2006}).
\btitle{Bayesian auxiliary variable models for binary and multinomial
regression}.
\bjournal{Bayesian Anal.}
\bvolume{1}
\bpages{145--168 (electronic)}.
\bid{mr={2227368}}
\bptok{imsref}%
\end{barticle}
%
\endbibitem

%b17 #&#
\bibitem[\protect\citeauthoryear{Holmes and Mallick}{2003}]{holmmal03}
%
\begin{barticle}[mr]
\bauthor{\bsnm{Holmes},~\bfnm{C.~C.}\binits{C.~C.}} \AND
\bauthor{\bsnm{Mallick},~\bfnm{B.~K.}\binits{B.~K.}}
(\byear{2003}).
\btitle{Generalized nonlinear modeling with multivariate free-knot regression
splines}.
\bjournal{J. Amer. Statist. Assoc.}
\bvolume{98}
\bpages{352--368}.
\bid{doi={10.1198/016214503000143}, issn={0162-1459}, mr={1995711}}
\bptok{imsref}%
\end{barticle}
%
\endbibitem

%b18 #&#
\bibitem[\protect\citeauthoryear{Kang et~al.}{2011}]{KangJASA2011}
%
\begin{barticle}[mr]
\bauthor{\bsnm{Kang},~\bfnm{Jian}\binits{J.}},
\bauthor{\bsnm{Johnson},~\bfnm{Timothy~D.}\binits{T.~D.}},
\bauthor{\bsnm{Nichols},~\bfnm{Thomas~E.}\binits{T.~E.}} \AND
\bauthor{\bsnm{Wager},~\bfnm{Tor~D.}\binits{T.~D.}}
(\byear{2011}).
\btitle{Meta analysis of functional neuroimaging data via {B}ayesian spatial
point processes}.
\bjournal{J. Amer. Statist. Assoc.}
\bvolume{106}
\bpages{124--134}.
\bid{doi={10.1198/jasa.2011.ap09735}, issn={0162-1459}, mr={2816707}}
\bptok{imsref}%
\end{barticle}
%
\endbibitem

%b19 #&#
\bibitem[\protect\citeauthoryear{Kober et~al.}{2008}]{Kober2008}
%
\begin{barticle}[pbm]
\bauthor{\bsnm{Kober},~\bfnm{Hedy}\binits{H.}},
\bauthor{\bsnm{Barrett},~\bfnm{Lisa~Feldman}\binits{L.~F.}},
\bauthor{\bsnm{Joseph},~\bfnm{Josh}\binits{J.}},
\bauthor{\bsnm{Bliss-Moreau},~\bfnm{Eliza}\binits{E.}},
\bauthor{\bsnm{Lindquist},~\bfnm{Kristen}\binits{K.}} \AND
\bauthor{\bsnm{Wager},~\bfnm{Tor~D.}\binits{T.~D.}}
(\byear{2008}).
\btitle{Functional grouping and cortical-subcortical interactions in
emotion: A
meta-analysis of neuroimaging studies}.
\bjournal{Neuroimage}
\bvolume{42}
\bpages{998--1031}.
\bid{doi={10.1016/j.neuroimage.2008.03.059}, issn={1095-9572},
mid={NIHMS118173}, pii={S1053-8119(08)00294-2}, pmcid={2752702},
pmid={18579414}}
\bptok{imsref}%
\end{barticle}
%
\endbibitem

%b20 #&#
\bibitem[\protect\citeauthoryear{Krivobokova, Crainiceanu and
Kauermann}{2008}]{KrivCraiKauefast2008}
%
\begin{barticle}[mr]
\bauthor{\bsnm{Krivobokova},~\bfnm{Tatyana}\binits{T.}},
\bauthor{\bsnm{Crainiceanu},~\bfnm{Ciprian~M.}\binits{C.~M.}} \AND
\bauthor{\bsnm{Kauermann},~\bfnm{G{\"o}ran}\binits{G.}}
(\byear{2008}).
\btitle{Fast adaptive penalized splines}.
\bjournal{J. Comput. Graph. Statist.}
\bvolume{17}
\bpages{1--20}.
\bid{doi={10.1198/106186008X287328}, issn={1061-8600}, mr={2424792}}
\bptok{imsref}%
\end{barticle}
%
\endbibitem

%b21 #&#
\bibitem[\protect\citeauthoryear{Lindquist}{2008}]{Lind2008}
%
\begin{barticle}[mr]
\bauthor{\bsnm{Lindquist},~\bfnm{Martin~A.}\binits{M.~A.}}
(\byear{2008}).
\btitle{The statistical analysis of f{MRI} data}.
\bjournal{Statist. Sci.}
\bvolume{23}
\bpages{439--464}.
\bid{doi={10.1214/09-STS282}, issn={0883-4237}, mr={2530545}}
\bptok{imsref}%
\end{barticle}
%
\endbibitem

%b22 #&#
\bibitem[\protect\citeauthoryear{Loader}{1999}]{load99}
%
\begin{bbook}[mr]
\bauthor{\bsnm{Loader},~\bfnm{Clive}\binits{C.}}
(\byear{1999}).
\btitle{Local Regression and Likelihood}.
\bpublisher{Springer}, \baddress{New York}.
\bid{mr={1704236}}
\bptok{imsref}%
\end{bbook}
%
\endbibitem

%b23 #&#
\bibitem[\protect\citeauthoryear{McCullagh and Nelder}{1989}]{Mccullagh1989}
%
\begin{bbook}[author]
\bauthor{\bsnm{McCullagh},~\bfnm{P.}\binits{P.}} \AND
\bauthor{\bsnm{Nelder},~\bfnm{J.}\binits{J.}}
(\byear{1989}).
\btitle{Generalized Linear Models}, \bedition{2}nd ed.
\bpublisher{Chapman \& Hall/CRC}, \baddress{Boca Raton, FL}.
\bptok{imsref}%
\end{bbook}
%
\endbibitem

%b24 #&#
\bibitem[\protect\citeauthoryear{O'Sullivan, Yandell and
Raynor}{1986}]{OSullivan86}
%
\begin{barticle}[mr]
\bauthor{\bsnm{O'Sullivan},~\bfnm{Finbarr}\binits{F.}},
\bauthor{\bsnm{Yandell},~\bfnm{Brian~S.}\binits{B.~S.}} \AND
\bauthor{\bsnm{Raynor},~\bfnm{William~J.}\binits{W.~J.} \bsuffix{Jr.}}
(\byear{1986}).
\btitle{Automatic smoothing of regression functions in generalized linear
models}.
\bjournal{J. Amer. Statist. Assoc.}
\bvolume{81}
\bpages{96--103}.
\bid{issn={0162-1459}, mr={0830570}}
\bptok{imsref}%
\end{barticle}
%
\endbibitem

%b25 #&#
\bibitem[\protect\citeauthoryear{Penny, Trujillo-Barreto and
Friston}{2005}]{penny05}
%
\begin{barticle}[pbm]
\bauthor{\bsnm{Penny},~\bfnm{William~D.}\binits{W.~D.}},
\bauthor{\bsnm{Trujillo-Barreto},~\bfnm{Nelson~J.}\binits{N.~J.}}
\AND
\bauthor{\bsnm{Friston},~\bfnm{Karl~J.}\binits{K.~J.}}
(\byear{2005}).
\btitle{Bayesian fMRI time series analysis with spatial priors}.
\bjournal{Neuroimage}
\bvolume{24}
\bpages{350--362}.
\bid{doi={10.1016/j.neuroimage.2004.08.034}, issn={1053-8119},
pii={S1053-8119(04)00497-5}, pmid={15627578}}
\bptok{imsref}%
\end{barticle}
%
\endbibitem

%b26 #&#
\bibitem[\protect\citeauthoryear{Psarakis and Panaretos}{1990}]{psarpanft90}
%
\begin{barticle}[mr]
\bauthor{\bsnm{Psarakis},~\bfnm{S.}\binits{S.}} \AND
\bauthor{\bsnm{Panaretos},~\bfnm{J.}\binits{J.}}
(\byear{1990}).
\btitle{The folded {$t$} distribution}.
\bjournal{Comm. Statist. Theory Methods}
\bvolume{19}
\bpages{2717--2734}.
\bid{doi={10.1080/03610929008830342}, issn={0361-0926}, mr={1086030}}
\bptok{imsref}%
\end{barticle}
%
\endbibitem

%b27 #&#
\bibitem[\protect\citeauthoryear{Rue and Held}{2005}]{GMRFbook}
%
\begin{bbook}[mr]
\bauthor{\bsnm{Rue},~\bfnm{H{\aa}vard}\binits{H.}} \AND
\bauthor{\bsnm{Held},~\bfnm{Leonhard}\binits{L.}}
(\byear{2005}).
\btitle{Gaussian {M}arkov Random Fields: Theory and Applications}.
\bseries{Monographs on Statistics and Applied Probability}
\bvolume{104}.
\bpublisher{Chapman \& Hall/CRC}, \baddress{Boca Raton, FL}.
\bid{doi={10.1201/9780203492024}, mr={2130347}}
\bptok{imsref}%
\end{bbook}
%
\endbibitem

%b28 #&#
\bibitem[\protect\citeauthoryear{Rue, Martino and Chopin}{2009}]{rueinla09}
%
\begin{barticle}[mr]
\bauthor{\bsnm{Rue},~\bfnm{H{\aa}vard}\binits{H.}},
\bauthor{\bsnm{Martino},~\bfnm{Sara}\binits{S.}} \AND
\bauthor{\bsnm{Chopin},~\bfnm{Nicolas}\binits{N.}}
(\byear{2009}).
\btitle{Approximate {B}ayesian inference for latent {G}aussian models
by using
integrated nested {L}aplace approximations}.
\bjournal{J. R. Stat. Soc. Ser. B Stat. Methodol.}
\bvolume{71}
\bpages{319--392}.
\bid{doi={10.1111/j.1467-9868.2008.00700.x}, issn={1369-7412}, mr={2649602}}
\bptnote{check related}%
\bptok{imsref}%
\end{barticle}
%
\endbibitem

%b29 #&#
\bibitem[\protect\citeauthoryear{Ruppert and
Carroll}{2000}]{RuppCarrspat2000}
%
\begin{barticle}[author]
\bauthor{\bsnm{Ruppert},~\bfnm{David}\binits{D.}} \AND
\bauthor{\bsnm{Carroll},~\bfnm{Raymond~J.}\binits{R.~J.}}
(\byear{2000}).
\btitle{Spatially-adaptive penalties for spline fitting}.
\bjournal{Aust. N. Z. J. Stat.}
\bvolume{42}
\bpages{205--223}.
\bptok{imsref}%
\end{barticle}
%
\endbibitem

%b30 #&#
\bibitem[\protect\citeauthoryear{Spiegelhalter et~al.}{2002}]{spiegdic02}
%
\begin{barticle}[mr]
\bauthor{\bsnm{Spiegelhalter},~\bfnm{David~J.}\binits{D.~J.}},
\bauthor{\bsnm{Best},~\bfnm{Nicola~G.}\binits{N.~G.}},
\bauthor{\bsnm{Carlin},~\bfnm{Bradley~P.}\binits{B.~P.}} \AND
\bauthor{\bparticle{van~der} \bsnm{Linde},~\bfnm{Angelika}\binits{A.}}
(\byear{2002}).
\btitle{Bayesian measures of model complexity and fit}.
\bjournal{J. R. Stat. Soc. Ser. B Stat. Methodol.}
\bvolume{64}
\bpages{583--639}.
\bid{doi={10.1111/1467-9868.00353}, issn={1369-7412}, mr={1979380}}
\bptnote{check related}%
\bptok{imsref}%
\end{barticle}
%
\endbibitem

%b31 #&#
\bibitem[\protect\citeauthoryear{Talairach and Tournoux}{1988}]{TalTour88}
%
\begin{bbook}[author]
\bauthor{\bsnm{Talairach},~\bfnm{J.}\binits{J.}} \AND
\bauthor{\bsnm{Tournoux},~\bfnm{P.}\binits{P.}}
(\byear{1988}).
\btitle{Co-planar Stereotaxic Atlas of the Human Brain: 3-Dimensional
Proportional System---an Approach to Cerebral Imaging}.
\bpublisher{Thieme Medical Publishers}, \baddress{New York}.
\bptok{imsref}%
\end{bbook}
%
\endbibitem

%b32 #&#
\bibitem[\protect\citeauthoryear{Tipping}{2001}]{tippspar2001}
%
\begin{barticle}[mr]
\bauthor{\bsnm{Tipping},~\bfnm{Michael~E.}\binits{M.~E.}}
(\byear{2001}).
\btitle{Sparse {B}ayesian learning and the relevance vector machine}.
\bjournal{J.~Mach. Learn. Res.}
\bvolume{1}
\bpages{211--244}.
\bid{doi={10.1162/15324430152748236}, issn={1532-4435}, mr={1875838}}
\bptok{imsref}%
\end{barticle}
%
\endbibitem

%b33 #&#
\bibitem[\protect\citeauthoryear{Trippa and Muliere}{2009}]{trimulspl09}
%
\begin{barticle}[mr]
\bauthor{\bsnm{Trippa},~\bfnm{Lorenzo}\binits{L.}} \AND
\bauthor{\bsnm{Muliere},~\bfnm{Pietro}\binits{P.}}
(\byear{2009}).
\btitle{Bayesian nonparametric binary regression via random tessellations}.
\bjournal{Statist. Probab. Lett.}
\bvolume{79}
\bpages{2273--2280}.
\bid{doi={10.1016/j.spl.2009.07.026}, issn={0167-7152}, mr={2591984}}
\bptok{imsref}%
\end{barticle}
%
\endbibitem

%b34 #&#
\bibitem[\protect\citeauthoryear{Turkeltaub et~al.}{2002}]{Turk2002}
%
\begin{barticle}[author]
\bauthor{\bsnm{Turkeltaub},~\bfnm{P.}\binits{P.}},
\bauthor{\bsnm{Eden},~\bfnm{G.}\binits{G.}},
\bauthor{\bsnm{Jones},~\bfnm{K.}\binits{K.}} \AND
\bauthor{\bsnm{Zeffiro},~\bfnm{T.~A.~T.}\binits{T.~A.~T.}}
(\byear{2002}).
\btitle{Meta-analysis of the functional neuroanatomy of single-word reading:
Method and validation}.
\bjournal{NeuroImage}
\bvolume{16}
\bpages{765--780}.
\bptok{imsref}%
\end{barticle}
%
\endbibitem

%b35 #&#
\bibitem[\protect\citeauthoryear{Wager, Jonides and
Reading}{2004}]{Wager2004}
%
\begin{barticle}[pbm]
\bauthor{\bsnm{Wager},~\bfnm{Tor~D.}\binits{T.~D.}},
\bauthor{\bsnm{Jonides},~\bfnm{John}\binits{J.}} \AND
\bauthor{\bsnm{Reading},~\bfnm{Susan}\binits{S.}}
(\byear{2004}).
\btitle{Neuroimaging studies of shifting attention: A meta-analysis}.
\bjournal{Neuroimage}
\bvolume{22}
\bpages{1679--1693}.
\bid{doi={10.1016/j.neuroimage.2004.03.052}, issn={1053-8119},
pii={S105381190400223X}, pmid={15275924}}
\bptok{imsref}%
\end{barticle}
%
\endbibitem

%b36 #&#
\bibitem[\protect\citeauthoryear{Wager, Lindquist and
Kaplan}{2007}]{WagerSCAN2007}
%
\begin{barticle}[author]
\bauthor{\bsnm{Wager},~\bfnm{Tor~D.}\binits{T.~D.}},
\bauthor{\bsnm{Lindquist},~\bfnm{Martin~A.}\binits{M.~A.}} \AND
\bauthor{\bsnm{Kaplan},~\bfnm{Lauren}\binits{L.}}
(\byear{2007}).
\btitle{Meta-analysis of functional neuroimaging data: Current and future
directions}.
\bjournal{Social Cognitive and Affective Neuroscience}
\bvolume{2}
\bpages{150--158}.
\bptok{imsref}%
\end{barticle}
%
\endbibitem

%b37 #&#
\bibitem[\protect\citeauthoryear{Wager et~al.}{2008}]{WagerHB2008}
%
\begin{bmisc}[author]
\bauthor{\bsnm{Wager},~\bfnm{T.~D.}\binits{T.~D.}},
\bauthor{\bsnm{Barrett},~\bfnm{L.~F.}\binits{L.~F.}},
\bauthor{\bsnm{Bliss-Moreau},~\bfnm{E.}\binits{E.}},
\bauthor{\bsnm{Lindquist},~\bfnm{K.}\binits{K.}},
\bauthor{\bsnm{Duncan},~\bfnm{S.}\binits{S.}},
\bauthor{\bsnm{Kober},~\bfnm{H.}\binits{H.}},
\bauthor{\bsnm{Joseph},~\bfnm{J.}\binits{J.}},
\bauthor{\bsnm{Davidson},~\bfnm{M.}\binits{M.}} \AND
\bauthor{\bsnm{Mize},~\bfnm{J.}\binits{J.}}
(\byear{2008}).
\bhowpublished{The neuroimaging of emotion.
In \textit{Handbook of Emotion}
({M.}~{Lewis}, ed.)
249--271.
Guilford Press, New York}.
\bptok{imsref}%
\end{bmisc}
%
\endbibitem

%b38 #&#
\bibitem[\protect\citeauthoryear{Wager et~al.}{2009}]{WagerNI2009}
%
\begin{barticle}[pbm]
\bauthor{\bsnm{Wager},~\bfnm{Tor~D.}\binits{T.~D.}},
\bauthor{\bsnm{Lindquist},~\bfnm{Martin~A.}\binits{M.~A.}},
\bauthor{\bsnm{Nichols},~\bfnm{Thomas~E.}\binits{T.~E.}},
\bauthor{\bsnm{Kober},~\bfnm{Hedy}\binits{H.}} \AND
\bauthor{\bsnm{Van Snellenberg},~\bfnm{Jared X.}\binits{J.~X.}}
(\byear{2009}).
\btitle{Evaluating the consistency and specificity of neuroimaging
data using
meta-analysis}.
\bjournal{Neuroimage}
\bvolume{45}
\bpages{S210--S221}.
\bid{doi={10.1016/j.neuroimage.2008.10.061}, issn={1095-9572},
pii={S1053-8119(08)01211-1}, pmid={19063980}}
\bptok{imsref}%
\end{barticle}
%
\endbibitem

%b39 #&#
\bibitem[\protect\citeauthoryear{Wahba et~al.}{1995}]{wahba97}
%
\begin{barticle}[mr]
\bauthor{\bsnm{Wahba},~\bfnm{Grace}\binits{G.}},
\bauthor{\bsnm{Wang},~\bfnm{Yuedong}\binits{Y.}},
\bauthor{\bsnm{Gu},~\bfnm{Chong}\binits{C.}},
\bauthor{\bsnm{Klein},~\bfnm{Ronald}\binits{R.}} \AND
\bauthor{\bsnm{Klein},~\bfnm{Barbara}\binits{B.}}
(\byear{1995}).
\btitle{Smoothing spline {ANOVA} for exponential families, with
application to
the {W}isconsin {E}pidemiological {S}tudy of {D}iabetic {R}etinopathy}.
\bjournal{Ann. Statist.}
\bvolume{23}
\bpages{1865--1895}.
\bid{doi={10.1214/aos/1034713638}, issn={0090-5364}, mr={1389856}}
\bptnote{check year}%
\bptok{imsref}%
\end{barticle}
%
\endbibitem

%b40 #&#
\bibitem[\protect\citeauthoryear{Wiper, Gir{\'o}n and
Pewsey}{2008}]{wipegiropewshnt08}
%
\begin{barticle}[mr]
\bauthor{\bsnm{Wiper},~\bfnm{M.~P.}\binits{M.~P.}},
\bauthor{\bsnm{Gir{\'o}n},~\bfnm{F.~J.}\binits{F.~J.}} \AND
\bauthor{\bsnm{Pewsey},~\bfnm{Arthur}\binits{A.}}
(\byear{2008}).
\btitle{Objective {B}ayesian inference for the half-normal and half-{$t$}
distributions}.
\bjournal{Comm. Statist. Theory Methods}
\bvolume{37}
\bpages{3165--3185}.
\bid{doi={10.1080/03610920802105184}, issn={0361-0926}, mr={2467759}}
\bptok{imsref}%
\end{barticle}
%
\endbibitem

%b41 #&#
\bibitem[\protect\citeauthoryear{Wood and Kohn}{1998}]{woodkohn98}
%
\begin{barticle}[author]
\bauthor{\bsnm{Wood},~\bfnm{Sally~A.}\binits{S.~A.}} \AND
\bauthor{\bsnm{Kohn},~\bfnm{Robert}\binits{R.}}
(\byear{1998}).
\btitle{A {B}ayesian approach to robust binary nonparametric regression}.
\bjournal{J. Amer. Statist. Assoc.}
\bvolume{93}
\bpages{203--213}.
\bptok{imsref}%
\end{barticle}
%
\endbibitem

%b42 #&#
\bibitem[\protect\citeauthoryear{Wood et~al.}{2008}]{woodkohn08}
%
\begin{barticle}[mr]
\bauthor{\bsnm{Wood},~\bfnm{Sally~A.}\binits{S.~A.}},
\bauthor{\bsnm{Kohn},~\bfnm{Robert}\binits{R.}},
\bauthor{\bsnm{Cottet},~\bfnm{Remy}\binits{R.}},
\bauthor{\bsnm{Jiang},~\bfnm{Wenxin}\binits{W.}} \AND
\bauthor{\bsnm{Tanner},~\bfnm{Martin}\binits{M.}}
(\byear{2008}).
\btitle{Locally adaptive nonparametric binary regression}.
\bjournal{J. Comput. Graph. Statist.}
\bvolume{17}
\bpages{352--372}.
\bid{doi={10.1198/106186008X318576}, issn={1061-8600}, mr={2439964}}
\bptok{imsref}%
\end{barticle}
%
\endbibitem

%b43 #&#
\bibitem[\protect\citeauthoryear{Yue, Loh and
Lindquist}{2010}]{yuelohlindfmri2009}
%
\begin{barticle}[mr]
\bauthor{\bsnm{Yue},~\bfnm{Yu}\binits{Y.}},
\bauthor{\bsnm{Loh},~\bfnm{Ji~Meng}\binits{J.~M.}} \AND
\bauthor{\bsnm{Lindquist},~\bfnm{Martin~A.}\binits{M.~A.}}
(\byear{2010}).
\btitle{Adaptive spatial smoothing of f{MRI} images}.
\bjournal{Stat. Interface}
\bvolume{3}
\bpages{3--13}.
\bid{issn={1938-7989}, mr={2609707}}
\bptok{imsref}%
\end{barticle}
%
\endbibitem

%b44 #&#
\bibitem[\protect\citeauthoryear{Yue and Loh}{2011}]{yuelohpp10}
%
\begin{barticle}[pbm]
\bauthor{\bsnm{Yue},~\bfnm{Yu~Ryan}\binits{Y.~R.}} \AND
\bauthor{\bsnm{Loh},~\bfnm{Ji~Meng}\binits{J.~M.}}
(\byear{2011}).
\btitle{Bayesian semiparametric intensity estimation for inhomogeneous spatial
point processes}.
\bjournal{Biometrics}
\bvolume{67}
\bpages{937--946}.
\bid{doi={10.1111/j.1541-0420.2010.01531.x}, issn={1541-0420}, pmid={21175553}}
\bptnote{check year}%
\bptok{imsref}%
\end{barticle}
%
\endbibitem

%b45 #&#
\bibitem[\protect\citeauthoryear{Yue and Speckman}{2010}]{yuespecknon09}
%
\begin{barticle}[mr]
\bauthor{\bsnm{Yue},~\bfnm{Yu}\binits{Y.}} \AND
\bauthor{\bsnm{Speckman},~\bfnm{Paul~L.}\binits{P.~L.}}
(\byear{2010}).
\btitle{Nonstationary spatial {G}aussian {M}arkov random fields}.
\bjournal{J.~Comput. Graph. Statist.}
\bvolume{19}
\bpages{96--116}.
\bid{doi={10.1198/jcgs.2009.08124}, issn={1061-8600}, mr={2654402}}
\bptok{imsref}%
\end{barticle}
%
\endbibitem

\end{thebibliography}
\end{document}